\documentclass[aps,twocolumn,english,notitlepage,pra]{revtex4-1}
\usepackage[T1]{fontenc}
\usepackage[latin9]{inputenc}
\usepackage{babel}
\usepackage{enumitem}
\usepackage{pifont}
\usepackage{verbatim}
\usepackage{float}
\usepackage{amsmath}
\usepackage{amsthm}
\usepackage{braket}
\usepackage{amssymb}
\usepackage{graphicx}
\usepackage{appendix}
\usepackage[dvipsnames]{xcolor}
\usepackage{array}
\usepackage{ulem}
\usepackage{subfiles}
\usepackage{multirow}
\usepackage{booktabs}
\usepackage{hyperref}
\hypersetup{
	colorlinks=true,
	linkcolor=blue,
	filecolor=magenta,      
	urlcolor=cyan,
	citecolor=red
}
\newcolumntype{P}[1]{>{\centering\arraybackslash}p{#1}}

\usepackage{subfigure,bbm}

\usepackage{mathtools}

\newcommand{\be}{\begin{equation}}
\newcommand{\ee}{\end{equation}}
\newcommand{\ba}{\begin{align}}
\newcommand{\ea}{\end{align}}

\newcommand{\red}[1]{\textcolor{black}{#1}}

\makeatletter

\begin{document}

\title{Multiplexed quantum repeaters \\
based on dual-species trapped-ion systems}

\author{Prajit Dhara$^{1}$}
\author{Norbert M. Linke$^{2}$}
\author{Edo Waks$^{3}$}
\author{Saikat Guha$^{1}$}
\author{Kaushik P. Seshadreesan$^{1,4}$}
\email{kausesh@pitt.edu}
\affiliation{$^{1}$Wyant College of Optical Sciences, University of Arizona, Tucson, AZ 85721, USA}
\affiliation{$^{2}$Joint Quantum Institute and Department of Physics, University of Maryland, College Park, MD 20742, USA}
\affiliation{$^{3}$Joint Quantum Institute and Department of Electrical and Computer Engineering, University of Maryland, College Park, MD 20742, USA}
\affiliation{$^{4}$Department of Informatics and Networked Systems, School of Computing and Information, University of Pittsburgh, Pittsburgh, PA 15260, USA}

\begin{abstract}
    Trapped ions form an advanced technology platform for quantum information processing with long qubit coherence times, high-fidelity quantum logic gates, optically active qubits, and a potential to scale up in size while preserving a high level of connectivity between qubits. 
    These traits make them attractive not only for quantum computing, but also for quantum networking. 
    Dedicated, special-purpose trapped-ion processors in conjunction with suitable interconnecting hardware can be used to form quantum repeaters that enable high-rate quantum communications between distant trapped-ion quantum computers in a network. 
    In this regard, hybrid traps with two distinct species of ions, where one ion species can generate ion-photon entanglement that is useful for optically interfacing with the network and the other has long memory lifetimes, useful for qubit storage, have been proposed for entanglement distribution. 
    We consider an architecture for a repeater based on such dual-species trapped-ion systems. 
    We propose and analyze a protocol based on spatial and temporal mode multiplexing for entanglement distribution across a line network of such repeaters. 
    Our protocol offers enhanced rates compared to rates previously reported for such repeaters. 
    We determine the ion resources required at the repeaters to attain the enhanced rates, and the best rates attainable when constraints are placed \red{on the number of repeaters and the number of ions per repeater.} 
    Our results bolster the case for near-term trapped ion systems as quantum repeaters for long distance quantum communications.
\end{abstract}

\date{\today}

\maketitle

\section{Introduction}
\label{intro}
Quantum information processing is set to revolutionize computation, communication and sensing technologies~\cite{Deutsch2020-yg,Dowling2003-uf}.
Applications of these technologies range from quantum speedups with NISQ processors~\cite{Bharti2021-es} to quantum key distribution~\cite{Xu2020-zr}, to quantum-enhanced distributed sensors~\cite{Zhang2020-py}. 
Quantum technologies are currently being actively developed across different physical platforms---from solid-state systems such as superconducting circuits~\cite{Kjaergaard2020-mq} and nitrogen vacancies~\cite{Pezzagna2021-zf} in diamond, to trapped ions~\cite{Bruzewicz2019-tu}, to nano-photonic systems~\cite{Flamini2019-me}. 
Quantum networks capable of faithfully transferring quantum states between nodes, including the capability to distribute quantum entanglement~\cite{Horodecki2009-lt}, are being developed both over short distances to scale up quantum computers in a modular fashion, as well as over long-distances to connect remote quantum computers, or a local area network of computers across physical platforms towards building a global quantum internet~\cite{Wehner2018-hg,Pirandola2016-kc,Kimble2008-yx}. 

Given that photons are the best transmitters of quantum information that can be used to implement scalable quantum communications, the primary challenge in quantum networking is the fundamental rate-loss trade-off. This trade-off exists for quantum communications over a lossy optical communication channel that models imperfections such as photon collection, coupling and detection inefficiencies, as well as transmission losses. 
The entanglement distribution capacity of the pure-loss optical channel with unlimited signal power and unlimited local quantum operations and classical communications (LOCC) is given by $C(\eta)=-\log_2(1-\eta)$ ebits per channel-use~\cite{pirandola2017a}, where $\eta$ is the channel transmissivity, and an ebit denotes a pair of maximally entangled qubits. 
In the limit of low transmissivity $\eta\ll 1$, this quantity scales as $\propto \eta$~\cite{Takeoka2014-ya}. 
As a result, in long-distance communications, say, over an optical fiber link whose transmissivity decreases exponentially with distance as $e^{-\alpha l}$, ($\alpha$ being the fiber loss coefficient per unit length), the entanglement distribution capacity drops exponentially with distance independent of the presence or absence of other imperfections. Quantum repeaters~\cite{Briegel1998-va,Munro2015-hv} help overcome this challenge. 
They are special-purpose quantum computers typically consisting of quantum sources, detectors, elementary logic gates and quantum memories. Quantum repeater architectures based on different physical platforms~\cite{Ghalaii2020-qc,Dias2020-lf,Seshadreesan2020-zl,Loock2020-wd,Sangouard2011-zm,Childress2005-xe} along successive generations of improved protocols~\cite{Muralidharan2016-wp,dhara2021} have been proposed that can achieve enhanced entanglement distribution rates beyond the direct transmission capacity.

Establishing a large-scale quantum network typically calls for setting up long-distance core networks. 
Among the large variety of physical systems that can be utilized to realize quantum repeaters for the core quantum network backbone, trapped-ion based systems form an excellent, robust choice, due to their inherently long memory coherence times~\cite{Monroe2007-us}. 
Moreover, trapped-ions are known to be an advanced qubit technology~\cite{Bruzewicz2019-tu}, and one of the front runners in the race for scalable universal quantum information processing~\cite{Monroe2013-no}. 
Repeater networks consisting of single-species trapped-ion nodes have been considered and analyzed in-depth in Ref.~\cite{Sangouard2009-zy}. 
More recently, Santra et al.~\cite{Santra2019-di} analyzed repeaters based on ion traps consisting of two species of ions with complementary properties---a \textit{communication ion} species with good optical properties that enables the network nodes to communicate with one another, and a \textit{memory ion} species having a long coherence time, and therefore suitable for information storage and efficient local quantum processing. 
Examples of such complementary pairs of ion species include $^{138}$Ba$^+$ and $^{171}$Yb$^+$, and $^{9}$Be$^+$ and $^{25}$Mg$^+$. 
In the former, e.g., the $^{138}$Ba$^+$ ion can emit a photon in the visible part of the spectrum ($493$ nm) that is entangled with the atomic state of the ion. 
Entanglement can be heralded between the atomic states of two such $^{138}$Ba$^+$ ions by performing an optical Bell-state measurement~\cite{Grice2011-hw} on the photons they emit. 
Such optically-mediated entanglement, when heralded between adjacent repeater nodes, can be faithfully transferred on to $^{171}$Yb$^+$ ions present at the respective nodes, whose atomic states have extremely long coherence times, thus allowing the storage of entanglement between the nodes which can later be processed using efficient quantum gate operations~\cite{Inlek2017-tp,Tan2015-nq}. 
Santra et al.~\cite{Santra2019-di} presented a repeater architecture based on such dual-species trapped ion (DSTI) modules, and discussed a set of logic gates needed to implement repeater protocols. 
They analyzed the quantum communication rates attainable over a line network of such DSTI repeaters using a multiplexed protocol.
The rates were shown to exceed those possible with direct transmission.

In the present article, we explore a protocol based on spatial and temporal multiplexing for the trapped-ion repeater architecture involving DSTI modules that is more general than the one considered in Ref.~\cite{Santra2019-di}.
In spatial multiplexing, multiple communication ions attempt to generate remote entanglement between every pair of adjacent repeater nodes at each time step of a well-defined clock cycle. 
In time multiplexing, remote entanglement is heralded from entanglement generation attempts across a block of multiple time steps. 
While both spatial and time multiplexing were also considered in Ref.~\cite{Santra2019-di}, the latter was only considered implicitly with a fixed clock cycle duration for the photon-ion entanglement generation at the repeater nodes determined by the distance between adjacent nodes. 
Here, we treat the clock cycle duration as a free parameter, so that ion-photon entanglement generation at the nodes can be attempted at rates independent of the internodal spacing. 
This enables higher quantum communication rates than the rates supported by the protocol of Ref.~\cite{Santra2019-di}. 
Our protocol warrants suitably larger number of communication ions and memory ions at the repeater nodes for ion-photon entanglement generation and for storing the unheralded ion qubits, respectively. 
We determine the enhanced rates enabled by such a general protocol for different spatial multiplexing, numerically optimized over the number of repeaters and temporal multiplexing. 
We identify \red{the number of repeaters required and} the number of communication and memory ions required per repeater for the optimal implementation of the protocol, and discuss how the rates deteriorate from their optimum values when \red{the number of repeaters, or} the number of ions per repeater is constrained. 

The article is organized as follows. 
In Sec.~\ref{section:architecture}, we present a general architecture for the trapped-ion repeaters, along with the node operations and an associated error model. 
We also summarize the different timing parameters of the repeaters here. 
In Sec.~\ref{section:multiplexing}, we outline the concepts of spatial and time multiplexing-based quantum repeater protocols. 
Section~\ref{sec: protocols} contains our proposed protocol based on spatial and time multiplexing for the DSTI repeaters for the case of $^{138}$Ba$^+$ and $^{171}$Yb$^+$ ions, along with numerical results. 
We conclude with a discussion and summary in Sec.~\ref{sec: discussion}.

\begin{figure*}
	\centering
	\includegraphics[width=0.9\textwidth]{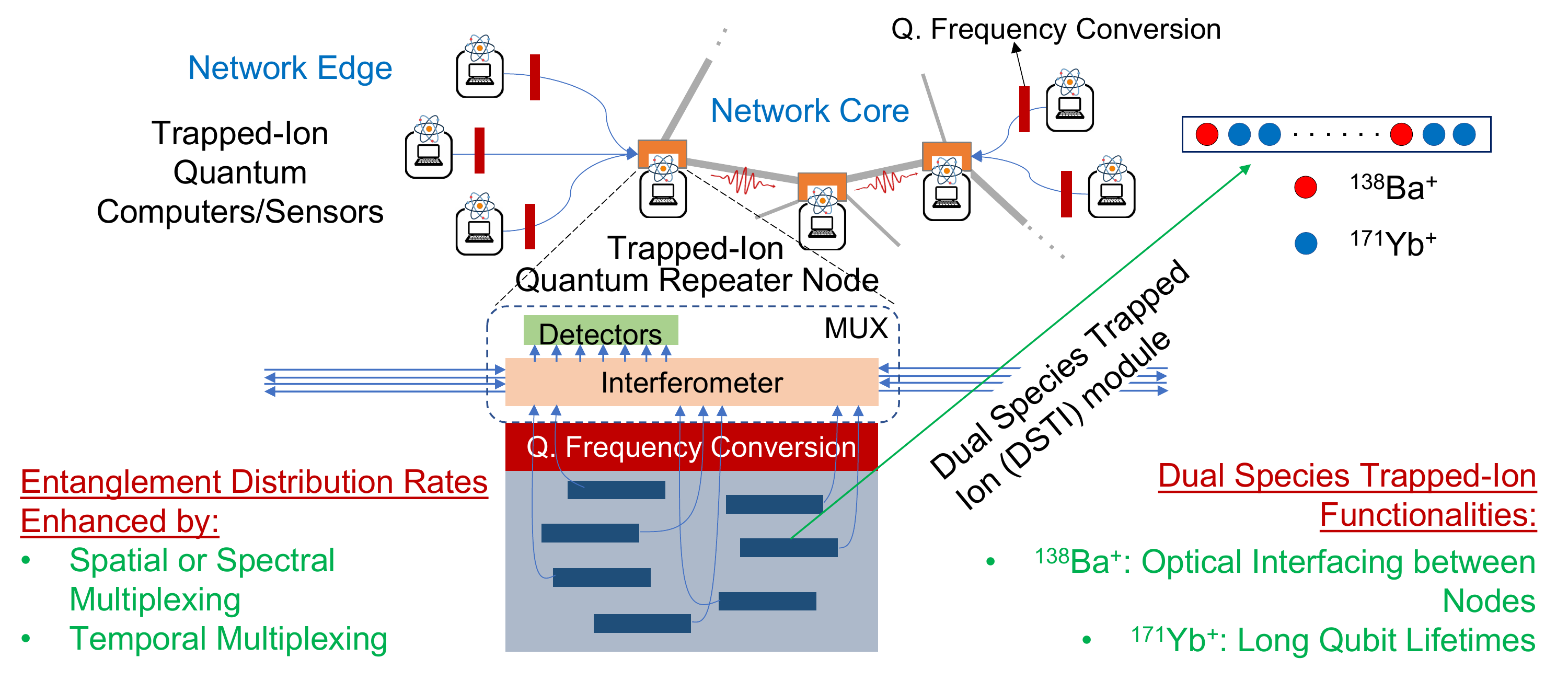}
	\caption{General architecture of a repeater node based on dual-species trapped ion (DSTI) modules to support entanglement distribution protocols based on mode multiplexing. The lines denote optical fibers.}
	\label{fig:node_arch}
\end{figure*}
\vspace{0.5 cm}

\section{Trapped-Ion Repeater Architecture}
\label{section:architecture}
The general architecture of the trapped-ion repeaters and the overall network analyzed in this work is depicted in Fig.~\ref{fig:node_arch}. 
\red{The repeaters consist of multiple DSTI modules containing i) $^{138}$Ba$^+$ ions, which serve as the communication ions, and ii) $^{171}$Yb$^+$ ions, which serve as the memory ions. 
The DSTI modules may thus be thought of as consisting of two independent ion ensembles. 
Each repeater node is equipped with lasers and light collection apparatuses for i) ion-photon entanglement generation using the communication ions, and for ii) performing qubit logic gates and measurement readouts on the memory ions. 
Since the qubits states in the two ion species have different transition frequencies, the above functions involve different lasers and thus do not affect each other.} 

\red{The photons emitted by the communication ions, upon collection, are first frequency converted to telecom wavelengths for inter-nodal transmissions~\cite{Hannegan2021-mm,Siverns2019-qf}, and then coupled into optical fibers. 
The communication ions in a DSTI module are assumed to be well spaced out so that the rate of resonant re-absorption of the photon emitted from one communication ion by another is low. 
Light collection is assumed to be spatially resolved so that light from different communication ions can be fiber coupled and transmitted over distinct spatial modes. 
The repeater nodes are assumed to be linked by fiber bundles capable of transmitting multiple single photons in distinct spatial modes to support spatial multiplexing.}

\red{The qubit logic gate operations on the memory ion qubits in a DSTI module are effected using highly collimated laser beams that address individual ions. 
However, for technical simplicity of operation of the DSTI modules, the measurement readouts of the memory ions in a module are considered to be global. 
This is because, readouts are effected by stimulating state-dependent fluorescence that can cause high levels of cross talk among the memory ions in the module even when individual ions are addressed for readout. 
It must be noted, however, that strategies to circumvent this problem have been successfully demonstrated, such as separating and shuttling selective ions into a separate zone for readout~\cite{Crain2019-zv}, which requires a complex trap geometry, or using another species for readout~\cite{Negnevitsky2018-ci}, which is challenging to combine with networking as it involves two species already.}

\red{Each repeater node also carries a quantum multiplexer (MUX)~\cite{Lee2020-nb, Krastanov2020-no}, whose functionalities include: i) optical switching, and ii) linear optical Bell-state measurements (BSMs). 
With photons transmitted between nodes in well-indexed spatial modes, independent Bell state measurements (discussed under repeater operations below) over the different multiplexed spatial modes can be effected using the MUX. 
The detectors are assumed to be ultrafast so that they can be reset and used over successive time steps. 
In other words, it is assumed that optical BSMs can be effected on successive time-bin modes using the same MUX to support temporal multiplexing. The nodes are assumed to share a common clock reference.}

\noindent {\it Repeater operations:} The basic repeater operations between and at the DSTI modules that we consider were proposed in Santra et al.~\cite{Santra2019-di}, and are summarized below. 
Interactions between DSTI modules, both within a repeater node as well as between adjacent repeater nodes are optically mediated. 
Photons emitted by the communication ions are duly collected, coupled into optical fiber, and interfered and measured to realize Bell state measurements. 
The simplest linear optical Bell state measurement for photonic qubits succeeds probabilistically. 
When a successful optical Bell state measurement is performed on photons from two ions, it results in entanglement being heralded between the ions. 
The atomic states of the communication ions are transferred or ``swapped'' to the memory ions  by ion-ion gates based on Coulomb interactions such as the Molmer-Sorensen gate~\cite{molmer_multiparticle_1999} to store entanglement over the long coherence times of the memory ions. 
The action of the swap gate $S_{c\rightarrow m}$ (where $c$ and $m$ labels denote the communication and memory ions, respectively) on a pair of entangled communication ions is given by
\begin{align}
\begin{split}
S_{c_1\rightarrow m_1}\otimes S_{c_2\rightarrow m_2}&|\psi\rangle_{m_1}|\beta\rangle_{c_1c_2}|\psi\rangle_{m_2}\\
& =|\psi\rangle_{c_1}|\beta\rangle_{m_1m_2}|\psi\rangle_{c_2}.
\end{split}
\end{align}
Here, the various quantum states are, $ \ket{\beta}_{c_1c_2}$ for the entangled communication ions, $\ket{\psi}_{m_i}$ for the memory ions before the linear optical entanglement swap, and $\ket{\psi}_{c_i}$ for the communication ions after the ion-ion swapping operation.

Finally, the entanglement swap operation between two entangled memory-ion pairs $|\beta\rangle_{m_1m_2}$ and $|\beta\rangle_{m_3m_4}$, when ions $m_2,m_3$ are in the same DSTI module, is accomplished by a CNOT gate operation followed by $Z$ and $X$ basis measurements, \red{where the latter may be effected using Hadamard gates followed by $Z$ basis measurements.} 
This operation extends the range of entanglement by establishing entanglement between $m_1$ and $m_4$. 
\red{With regard to implementing the measurements as part of the entanglement swap operations, firstly the gates constituting the entanglement swaps (CNOTs and Hadamards) are applied on individual memory ions in a DSTI module. 
Subsequently, the $Z$ basis measurements, as mentioned earlier, are effected via a global measurement readout of all the memory ions in the module. 
This forms an important consideration in our repeater protocol, wherein all the memory ions in a DSTI modules end up being measured in the $Z$ basis periodically as dictated by the protocol.}
\vspace{0.5 cm}

\noindent {\it Error model:} The success probability of optically-mediated heralded entanglement generation between communication ions present at two adjacent repeater nodes is given by $p=\frac{1}{2}\eta_c^2\eta_d^2e^{-\alpha L_0}$. 
Here, $\eta_c$ is the collection and coupling efficiency for the optical elements, $\eta_d$ is the efficiency of the detectors used in the Bell state measurement circuit, $\alpha$ is the fiber attenuation parameter, which is typically 0.2 dB/km at 1550 nm, and $L_0$ is the inter-repeater spacing
(detector dark counts and frequency conversion inefficiencies are neglected in the present analysis, and will be considered in future works).
When the communication ions are present at the same repeater node, but in different DSTI modules, the success probability is given by $p'=\frac{1}{2}\eta_c^2\eta_d^2,$ where it is assumed that the losses in transmission are negligible. 
Moreover, the entangled state of the communication ions is generally modeled by a Werner state of fidelity parameter $F_0$, given by
\begin{align}
\rho_{c_1,c_2}=F_0 \Phi^++\frac{1-F_0}{3}\left(\Phi^-+\Psi^++\Psi^-\right),
\label{cc_Werner_state}
\end{align}
where $\Phi^\pm=|\Phi^\pm\rangle\!\langle\Phi^\pm|$ and $\Psi^\pm=|\Psi^\pm\rangle\!\langle\Psi^\pm|$ are maximally entangled qubit Bell state density operators, with $|\Phi^\pm\rangle=\left(|0,1\rangle\pm|1,0\rangle\right)/2$ and $|\Psi^\pm\rangle=\left(|0,0\rangle\pm|1,1\rangle\right)/2$, and $\{|0\rangle,|1\rangle\}$ being the computational $Z$ basis eigenstates of the qubits. 
The Werner state model accounts for errors in the communication ions that may be caused by the presence of dephasing noise in the photonic qubits that undergo optical Bell state measurement.

Errors in the swap gate are compactly modeled jointly for a pair of instances of swap gates acting on two entangled communication ions to store the entanglement in two memory ions. The model is a two-qubit Pauli channel acting on the initial entangled state of the two communication ions resulting in a noisy mapping onto two memory ions (see Santra et al.~\cite[Eq. 3]{Santra2019-di} for details), and is described as
\begin{align}
\rho_{m_1m_2}&=(1-\epsilon_g)\rho_{c_1c_2}+\frac{\epsilon_g}{16}\sum_{k,k'=0}^3\sigma_{k'}\otimes \sigma_k\rho_{c_1c_2}\sigma_{k'}\otimes \sigma_k,\\
&=F_i\Phi^++\frac{1-F_i}{3}\left(\Phi^-+\Psi^++\Psi^-\right),\\
F_i&=(1-\epsilon_g)F_0+\frac{\epsilon_g}{4},
\end{align}
where \red{$\{\sigma_k\}_{k=0,1,2,3}$ are the Pauli matrices $I,X,Y$ and $Z$,} the state $\rho_{c_1c_2}$ is the Werner state of (\ref{cc_Werner_state}) and $\epsilon_g$ is the error parameter associated with the swap gate.

Finally, imperfections in the entanglement swap operation are modeled in two parts, namely, errors associated with i) the CNOT gate, and ii) the $X,Z$ measurement. 
\noindent i) The action of a noisy CNOT gate, also of error parameter $\epsilon_g$, acting on two qubits $m_2,m_3$ initially in state $\rho_{m_2}\otimes \rho_{m_3}$, is modeled as
\begin{align}
\rho'_{m_2m_3}&=(1-\epsilon_g)\textrm{CNOT}\left(\rho_{m_2}\otimes \rho_{m_3}\right)\textrm{CNOT}^\dagger\nonumber\\
&+\epsilon_g\frac{I_{m_2m_3}}{4},
\end{align}
where $m_2,m_3$ are the control and the target qubits, respectively, for the CNOT operation and $I_{m_2 m_3}$ is the two-qubit identity operator. 
Note that $\epsilon_g$ is an overestimate for the error in this gate, since it involves fewer Coulomb gates than the swap operation. 
\noindent ii) Errors associated with the $X$ and $Z$ measurements on the control and target qubits of the CNOT gate are functions of the gate error parameter $\epsilon_g$ and the initial fidelity of the entangled Werner state of two communication ions. 
When entanglement swaps are performed across a chain of $n\in\mathbb{Z}^+$ repeater nodes, the final noisy entangled state heralded between memory ions at the end nodes of the chain can be described also as a Werner state of the form in (\ref{cc_Werner_state}) with a fidelity parameter given by $F_f=1-\frac{3}{2}Q(n)$, where
\begin{align}
Q(n)=\frac{1}{2}\left(1-\left(1-2\epsilon_g-\frac{4}{3}(1-F_0)\right)^n\right).\label{eqn: end-to-end_Q}
\end{align}

\vspace{0.5 cm}

\noindent {\it Timing Parameters:} The proposed repeaters have a few characteristic timing parameters that are summarized in Table~\ref{tab:timing_parameters}. 
Firstly, the clock cycle duration $\tau$ (or sometimes denoted as $\tau_{\textrm{rep}}$) is the primary time unit, which denotes the rate at which the repeater nodes attempt ion-photon entanglement generation. This is a tunable parameter for the repeater operation. 
Secondly, there are the gate and measurement times $\tau_g$. Typical values for this time are of the order of microseconds. For example, high fidelity swap from $^{138}$Ba$^+$ to $^{171}$Yb$^+$ has been demonstrated in 100$\mu$s~\cite{Inlek2017-tp, Tan2015-nq}. 
Thirdly, there are the lifetimes of the communication and memory ions $\tau_o,\tau_m$. A typical value for the former is $100\mu$s (as has been reported for $^{138}$Ba$^+$ ions), while the latter can be taken to be long---the lifetime of $^{171}$Yb$^+$ transitions have been engineered to run in the order of minutes~\cite{wang_single-qubit_2017}. The parameters $\tau_o,\tau_m,\tau_g$ are governed by the choice of ions and gate implementation and hence take fixed values in any given physical realization of the repeaters.

\begin{table}[ht]
	\centering
	\begin{tabular}{c l}
		\textrm{Timing Parameter} & \textrm{Associated Meaning} \\
		\toprule
		$\tau$ & Clock cycle duration \\
		$\tau_g$ & Ion-ion gate and measurement times \\
		$\tau_o$ & Communication ion lifetime \\
		$\tau_m$ & Memory ion lifetime \\ 
		\bottomrule
	\end{tabular}
	\caption{Timing parameters associated with trapped-ion repeaters.}
	\label{tab:timing_parameters}
\end{table}

\section{Repeater Protocols based on Spatial and temporal Multiplexing}
\label{section:multiplexing}
In this section, we provide a brief background on multiplexing-based repeater protocols.  To begin with, due to the no-cloning theorem~\cite{Wootters1982-ph}, unlike classical communication, the simple strategy of `amplify and re-transmit' is not physically viable for entanglement generation between two remote parties Alice and Bob. The rates for direct transmission of qubits over a $M\in\mathbb{Z}^+$ quantum channels with a source repetition rate of $1/\tau$ are limited by the repeaterless bound on the entanglement generation capacity, which for pure loss channels of transmissivity $\eta$, is given by~\cite{pirandola2017a}
\begin{align}
C_{\text{direct}}(\eta,M,\tau)= -\frac{M}{\tau}\log_2(1-\eta)\ \textrm{ebits/s},\label{PLOB_ebits/sec}
\end{align}
referred to as the PLOB bound hereinafter. The PLOB bound tends to be $ \propto\eta $ for $ \eta\ll 1$.  

There are multiple paradigmatic approaches using quantum repeaters to beat this bound.
A widely recognized classification of these approaches is in terms of the so-called \textit{one-way} versus \textit{two-way} repeaters.
\textit{One-way quantum repeaters} encode the transmitted qubits using error correcting codes and the task for quantum repeaters is to decode, correct for transmission errors, re-encode and transmit from pre-determined locations on the channel. 
Entanglement can be distributed using these repeaters by encoding and transmitting one share of a logically-encoded ebit through the repeater links. This is a similar strategy as repeaters for one way classical communication.
\textit{Two-way quantum repeaters}, on the other hand, rely on the generation of local entanglement on smaller segments of the network. These locally shared ebits are concatenated with the aid of entanglement swaps to eventually achieve shared entanglement between the end parties on the channel. Such protocols could potentially be interspersed with entanglement purification to improve the quality of the shared ebit that is ultimately generated. 

In this work, we focus on two-way repeaters. 
Two-way repeaters are equipped with sources of photonic entangled pairs, quantum memory (QM) registers (trapped memory ions for our proposed designs in this work) and additional circuitry to perform quantum logic on the qubits stored in the QM register (including entanglement swaps on the memory ions). We begin with the assumption that the sources can produce perfect Bell pairs on demand every $\tau$ seconds. For simplicity, we initially assume arbitrarily large QM registers and infinitely long coherence times for the qubits stored in the QM. This is a necessary consideration for a constraint-free analysis of multiplexing. 

For the simplest network topology, namely a line network connecting two communicating parties, the total link distance $L$ is divided into $n+1$ elementary links. The repeater stations occupy the nodes at either end of each elementary link in this segmented network. The core strategy of the protocol is to generate shared entanglement on the elementary links before attempting entanglement swaps (between QMs) internally in the repeater stations. 
This is achieved by performing a linear optical Bell state measurement (BSM) between the transmitted qubits from neighbouring repeater stations. The simplest linear optical BSM is a probabilistic operation, which has a probability $p\leq 1/2$ of succeeding. Note that $p$ is dependent on the length of the elementary links, i.e. increasing the length of the elementary links deteriorates $p$. Since loss on fiber scales exponentially with distance, $p\propto \exp(-\alpha L/(n+1))$. Note that increasing $n$, i.e., reducing the elementary link length boosts $p$. However, this also means that a larger number of elementary links must simultaneously succeed, which shows a clear trade-off. Futher we assume that the QM entanglement swap can succeed with a probability $q\leq1$, where $q=1$ is possible with high fidelity entanglement swapping gates for trapped ion qubits. The achievable rate is given by,
\begin{align}
R_0(n)= \frac{p^{n+1}\times q^n}{\tau}.
\label{eqn:no_mux}
\end{align}
Since $R_0(n)<e^{-\alpha L}/\tau$, we perform worse than with direct transmission.

However, with the aid of multiple parallel attempts, i.e., multiplexing, the entanglement generation rate can be engineered to surpass direct transmission. This is the natural strategy to consider when individual links can only be generated in a probabilistic manner; instead of independent single attempts succeeding simultaneously, we perform multiple parallel attempts for each elementary link and concatenate successful links. Spatial (or equivalently spectral) multiplexing is the easiest modification to this protocol which is based on this paradigm. Here, the design incorporates parallel channels spatially, i.e., with separate optical fibers. With this modification, instead of a single BSM attempt every time slot, we can perform $M$ attempts and look for one success. With a spatial multiplexing size of $M$, and by considering the
probability of at least one success per elementary link, the end-to-end entanglement generation rate is now given by,
\begin{align}
R_1(n)=\frac{(1-(1-p)^M)^{n+1}\times q^n}{\tau}.
\label{eqn:spatial_mux}
\end{align}
It has been shown that with optimal choice of $n$ and suitable $M$, the rate equation in Eqn.~\eqref{eqn:spatial_mux} can surpass the direct transmission PLOB bound at a given link length. In fact, the rate envelope for Eqn.~\eqref{eqn:spatial_mux} has been derived in \cite{guha2015}, and has been shown to scale as $R_1\propto e^{-s\alpha L}$ with $s<1$, which allows the protocol to surpass rates possible with direct transmission.

Another strategy for multiplexing is to accumulate successes from $m\in\mathbb{Z}^+$ attempts over blocks of the fundamental time slot of $\tau$ seconds. This is called time multiplexing and mimics the effect of using multiple channels without the necessity for additional physical channels. 
Hence, we can perform the entanglement swap between different QMs at a repeater node only after every $m$ time slots. An achievable entanglement generation rate for the time multiplexed protocol is given by
\begin{align}
R_2(n,m)= \frac{(1-(1-p)^m)^{n+1}\times q^n}{m\tau}.
\label{eqn:time_mux}
\end{align}
It has been shown in \cite{razavi2009,dhara2021} that a time-multiplexed protocol can achieve a sub-exponential rate-vs.-distance scaling i.e.\ $R_2\propto e^{-t\sqrt{\alpha L}}$ with $t<1$. This is an improved performance over spatial multiplexing, and it has been shown that a protocol may surpass the PLOB bound with just time multiplexing. However, in a practical implementation, time multiplexing requires highly reliable QMs and large switching trees that scale as $\log_2(m)$. Imperfections in these components can lead to the loss of the sub-exponential advantage~\cite{dhara2021}. In general, with the incorporation of both spatial and temporal multiplexing, a two-way repeater protocol can achieve the rate $R(L,n,m)$ given as
\begin{align}
R(L,n,m)=\frac{(1-(1-p)^{mM})^{n+1}\times q^n}{m\tau}.
\label{mMRate}
\end{align}

It is important to note that there is a key difference between the multiplexing degrees in the spatial $(M)$ and time $(m)$ strategies. Increasing $M$ in a spatially multiplexed protocol requires the use of additional channels, which may be highly constrained (i.e. we may be limited by the number of physical optical fibers). In fact it is generally something that the network architect cannot modify, and hence it is not practical to optimize the rate with respect to $M$. Rather, given a certain maximum value of $M$, the rate envelope, as derived in Ref.~\cite{guha2015}, gives us an idea about the viability of the protocol to surpass the PLOB bound. Increasing the time multiplexing degree $m$ is only governed by the lifetime of the QM. As long as the lifetime surpasses a certain threshold governed by the protocol design, we can modify $m$ without the need for additional resources. Unlike spatial multiplexing,  time multiplexing can boost the probability of link creation on the elementary link seemingly arbitrarily, by increasing $m$. However, by increasing $m$, the effective time step increases from $\tau$ to $m \tau$ which degrades the rate (see Eqn.~(\ref{eqn:time_mux})). The boost in the success probability of the link, along with the optimization of the number of quantum repeater (QR) nodes $n$, overcompensates degradation, and an optimal value of $m$ for a given $L$ achieves the sub-exponential scaling.

Note that the repeater protocol considered in this work belongs to the second generation in the classification of successive generations of repeater protocols~\cite{sreraman2016}. 
This is so, because they do not include intermediate, iterative entanglement distillation steps, \red{but instead are based on multiple redundant entanglement generation attempts across the elementary links over spatial or spectral, and temporal modes, which can be perceived as the use of a repetition-based error correcting code for elementary entanglement generation. 
This is followed by identifying the latest successful heralded elementary links across each time multiplexing block and synchronous entanglement swapping of these elementary links at the repeater nodes.}

\section{Multiplexing-based protocol for trapped-ion repeaters: Protocol Design and Evaluation}
\label{sec: protocols}
In this section, we present our proposed protocol based on spatial and time multiplexing for entanglement distribution across a line network of trapped-ion repeaters \red{described in Sec.~\ref{section:architecture}, followed by numerical performance analyses. 
The protocol design is independent of the number of repeater nodes, and it is assumed that the ion resource parameters of the repeater nodes are unlimited and can be chosen to be as large as necessary to support any choice of values of the clock cycle duration $\tau$ and multiplexing parameters $M,m$. 
However, when analyzing the performance of the protocol, we will also consider constraints on these resources since in practice they are often constrained.}

The unit distance (inter-repeater spacing) and multiplexing parameters used in defining our protocol, and the repeater resource parameters are listed in Tables~\ref{tab:Network_parameters} and \ref{tab:Resource_parameters}, respectively. 
For simplicity, similar to Ref.~\cite{Santra2019-di}, we will consider the case of one DSTI module per node, i.e., $s=1$. 
However, the protocol leverages the multiple communication ions present within the DSTI modules for multiplexed entanglement generation attempts across elementary links in the network. 
The fundamental time step $\tau$, or in other words, the clock cycle duration, multiples of which are used as time multiplexing blocks, is chosen as a free parameter, and not tied to the physical distance between the repeaters. 
This makes our protocol more general than the one presented in Ref.~\cite{Santra2019-di}.

\red{\subsection{Protocol Design and Rates}}

For $n$ equally spaced repeaters and a total distance $L$ (between the end nodes), consider a $(m, M)$ repeater protocol with spatial multiplexing $M\in\mathbb{Z}^+$ and time multiplexing $m\in\mathbb{Z}^+$. 
The inter-repeater spacing is given by $L_0=L/(n+1)$. 
For a given $L_0$, the time it takes for the heralding information of success or failure of optically-mediated entanglement generation across adjacent repeater nodes to arrive at the nodes is $T=L_0/c$, where $c$ is the speed of light in the optical fiber used for inter-repeater node transmissions (henceforth referred to as the heralding time). 
The protocol aims to successfully herald one elementary link entanglement in each elementary link from $m\times M$ total attempts spread over $m\tau$ seconds. 
The heralding time and the gates and measurement time together add up to dictate the rate of generating the elementary link entanglements.  
Since all the memory ions are in one DSTI module, entanglement swapping across these elementary link entangled memory ions, which can performed deterministically using CNOT gate followed by $X$ and $Z$ measurement, distributes entanglement between the end nodes.

\vspace{0.5 cm}

\noindent {\it Rate Formulas under ideal repeater operations.} Assuming gate operations at the repeaters to be ideal and the optical fibers to be pure loss channels (no dephasing errors) for the moment, the rate in ebits per second attained by the protocol is given by the general formula
\begin{align}
R=\frac{\left(1-(1-p)^{Mm}\right)^{n+1}}{\mathbf{T}},\label{eqn: generic_rate_s=1}
\end{align}
where the numerator denotes the probability of successfully heralding at least one entangled ion-ion pair across each of the $n+1$ elementary links ($p$ being the success probability of optical Bell swap discussed in Sec.~\ref{section:architecture}), and the $\mathbf{T}$ in the denominator is the time it takes to complete $m$ time steps of entanglement generation attempts across the elementary links. 
In order to attain optimal rates at any distance $L$, an optimal number of repeaters $n_{\textrm{opt}}$ would be required to be placed along the distance. 
Too few repeaters would result in excessive errors due to photon loss, whereas too many repeaters would result in excessive operational errors at the repeater nodes.
\begin{table}[ht]
	\centering
	\begin{tabular}{c l}
		\textrm{Parameter} & \textrm{Associated Meaning}\\
		\toprule
		$L_0$ & Inter-repeater spacing \\
		$M$ & Degree of spatial multiplexing \\
		$m$ & Degree of time multiplexing \\ 
		\bottomrule
	\end{tabular}
	\caption{Unit distance and multiplexing parameters.}
	\label{tab:Network_parameters}
\end{table}
\begin{table}[ht]
	\centering
	\begin{tabular}{c l}
		\textrm{Parameter} & \textrm{Associated Meaning} \\
		\toprule
		$s$ & DSTI modules per repeater \\
		$N_o$ & $^{138}$Ba$^+$ ions per DSTI module \\
		$N_m$ & $^{171}$Yb$^+$ ions per DSTI module \\ 
		\bottomrule
	\end{tabular}
	\caption{Resource parameters.}
	\label{tab:Resource_parameters}
\end{table}

Notice that the rate in Eqn.~\eqref{eqn: generic_rate_s=1} is a function of the parameters $m,M,$ and n along with physical system parameters such as collection and detection efficiencies $\eta_c, \eta_d$ and the total distance $L$ that enter the formula through $p=\frac{1}{2}\eta_c^2\eta_d^2e^{-\alpha L_0}$, where $L_0=L/(n+1)$. 
The denominator $\mathbf{T}$ is a function of the time multiplexing block length $m$ and the clock cycle duration $\tau$, but also depends on the ion-ion gate and measurement times $\tau_g$ and the heralding time $T$, which is in turn a function of $L_0$. The dependence on $\tau_g$ is due to the fact that it takes a  non-zero amount of time to perform the essential entanglement swap operations at the repeater nodes, which is $2\tau_g$ seconds ($\tau_g$ for the CNOT gate and $\tau_g$ for the $X,Z$ measurements). 
\begin{table*}
	\centering
	\renewcommand{\arraystretch}{1.25}
	\begin{tabular}{c c c c c c }
		\toprule[1.5pt]
		\parbox[t]{6mm}{\multirow{3}{*}{\rotatebox[origin=c]{90}{\kern-0.5em \textbf{Table A}}}} &  \multicolumn{2}{c}{Criterion: $T\geq\tau_o>\tau_g$} & Required $N_o$ & Required $N_m$ & Rate \\
		\cmidrule(lr){2-6}
		& & & \multirow{2}{*}{$2Mj$} & \multirow{2}{*}{$\leq 2Mm$} & \multirow{2}{*}{$\frac{\left(1-(1-p)^{Mm}\right)^{n+1}}{(k+m+2j-1)\tau}$} \\ \\
		\midrule[0.5pt]
		\midrule[0.5pt]
		\parbox[t]{6mm}{\multirow{3}{*}{\rotatebox[origin=c]{90}{\kern-1em \textbf{Table B}}}} &  \multicolumn{2}{c}{Criterion: $\tau_o>T\geq\tau_g$} & Required $N_o$ & Required $N_m$ & Rate \\
		\cmidrule(lr){2-6}
		& \textbf{Case 1}& $T+\tau_g>\tau_o>T$ & \multicolumn{3}{c}{\centering Same as Table A} \\ 
		\cmidrule(lr){2-6}
		& \textbf{Case 2}& $\tau_o\geq T+\tau_g>T$ & $2(Mk+j)$ & $2m$ & $\frac{\left(1-(1-p)^{Mm}\right)^{n+1}}{(k+m+3j-1)\tau}$ \\ 
		\midrule
		\midrule
		\parbox[t]{6mm}{\multirow{3}{*}{\rotatebox[origin=c]{90}{\kern-1em \textbf{Table C}}}}  & \multicolumn{2}{c}{Criterion: $\tau_o>\tau_g>T$} & Required $N_o$ & Required $N_m$ & Rate \\
		\cmidrule(lr){2-6}
		& \textbf{Case 1} & $T+\tau_g>\tau_o>\tau_g$ & 
		$2Mj$ &$\leq 2Mm$ &$\frac{\left(1-(1-p)^{Mm}\right)^{n+1}}{(m+3j-1)\tau}$\\ 
		\cmidrule(lr){2-6}
		& \textbf{Case 2} & $\tau_o\geq T+\tau_g>\tau_g$ & \multicolumn{3}{c}{\centering Same as Table B, Case 2} \\
		\bottomrule
	\end{tabular}
	\caption{Rates and ion requirements for $s=1$ operation of the $(m,M)$ multiplexed repeater protocol. The rate expressions correspond to ideal gate operations, and the optical fibers are assumed to be a pure loss channel. \red{For realistic gate operations, the overall rate is modified by the reverse coherent information of the end-to-end state, which is a function of the gate error parameter $\epsilon_g$ and the initial fidelity of the entangled Werner state $F_0$.} See Appendix~\ref{appendix:timing} for detailed timing charts and corresponding timing diagrams.}
	\label{tab:Rates_Ion_Requirements}
\end{table*}

The precise formula for the rate attainable with an $(m, M)$ repeater protocol over $n$ repeaters placed along a total distance $L$, along with the ion requirements to support the protocol are tabulated in Table~\ref{tab:Rates_Ion_Requirements}. 
The rate depends on the clock cycle duration for ion-photon entanglement generation attempts at the nodes $\tau$, and the relative values of the heralding time $T$, ion-ion gate times $\tau_g$ and the communication ion lifetime $\tau_o$. 
The value $\tau$ is allowed to be chosen independently of the heralding time, which distinguishes our protocol from the one presented in Ref.~\cite{Santra2019-di}. 
Consider $T=k\tau$ and the $\tau_g=j\tau$, where  and $j,k\in\mathbb{R}^+$. 
Values of $k,j<1$ clearly lead to sub-optimal rates, since they imply ample idle time at the repeater nodes, when the ions are not attempting entanglement distribution. 
Thus, we focus on operations that correspond to $j,k\geq1$. 
For simplicity of analysis, let us consider $j,k\in\mathbb{Z}^+$.
Values of $j=\tau_g/\tau>1$ can in principle be realized with multi-sector traps. 
These are traps with distinct sectors of ions (say $j$ of them) with each containing a batch of communication ions, such that at every time step, ions in one of the sectors are collectively excited, and the different sectors being excited in a cyclical fashion. 
It is implicitly and reasonably assumed that i) $\tau_o>\tau_g$ so that a communication ion's quantum state can be faithfully transferred to a memory ion with ion-ion gates before it irrecoverably decoheres, and ii) $\tau_{m}\gg m\tau$ for a large range of values $m$, so that the memory ions can be considered to be noise free. 
Among the $6= {3\choose{2}}$ orderings of the relative values of $T, \tau_g$ and $\tau_o$, due to the reasonable assumption $\tau_o>\tau_g$, we are left with 3 possible orderings, namely: $T\geq\tau_o>\tau_g$, $\tau_o>T\geq\tau_g$ and $\tau_o>\tau_g>T$. Table~\ref{tab:Rates_Ion_Requirements} discusses the rates and the ion requirements achieved by the repeater protocol for each of these cases. Timing charts and timing diagrams that describe the protocol including the operations at the repeater nodes from time-step to time-step, under these different conditions are elucidated in Appendix~\ref{appendix:timing}. 

As an example, consider the case $ T\geq\tau_0>\tau_g $ described in Table~\ref{tab:Table_A} in the Appendix. 
In this scenario, at every time step, $2M$ communication ions generate ion-photon entanglement, with $M$ of the photons being directed towards the left of the node and the other $M$ towards the right of the node. 
The moment these photons are generated, an ion-ion gate is initiated on each of the communication ions, to swap their atomic state into memory ions. 
For gate time $\tau_g=j\tau, j\in\mathbb{Z}^+$, at time $t=j\tau,$ the communication ions that were used to generate ion-photon entanglement at time step $t=0$ are freed up due to the completion of the ion-ion gate, and hence are ready to be reused. 
At this point, the first $2M$ atomic states have been loaded into memory ions. 
At time step $t=k\tau, k\in\mathbb{Z}^+$, the information about which two (one to the left of the node and one to the right), if any, of the $2M$ entanglement generation attempts at time $t=0$ actually heralded an elementary link entanglement, is received, at which point, the other $2(M-1)$ memory ions are freed up and ready for reuse. 
At time $t=(k+m-1)\tau$, similarly, all potentially successfully heralded elementary link entanglements across the time multiplexing block length of $m$ are stored in the memory ions. At this point, the repeater nodes choose the latest successful heralded link to the left and to the right and perform entanglement swap on those corresponding memory ions. Performing the entanglement swap involves measuring these memory ions, which takes a time duration $2\tau_g$, i.e., $2j\tau$. Thus, the rate  of distributing 1 ebit across the end nodes of the trapped-ion repeater chain is $\propto 1/(k+m-1+2j)\tau$. 
Since we consider global measurements that measure all ions in the DSTI, all the other accumulated entanglement resources at the nodes are also cleared in the process. 
The protocol then starts once again from time step $t=0$.

Note that typically with time-multiplexed repeaters the heralding time only causes latency in the protocol without affecting the rates~\cite{dhara2021}. However, in the present scheme of trapped-ion repeaters, as mentioned above, ion measurements are considered to be global, full-trap measurements that measure all ions present in a DSTI, as opposed to measurement of individual ions in a trap. 
As a result, all the other accumulated entangled resources at the nodes are also cleared in the process, which negatively impacts the entanglement distribution rates.

\vspace{0.5 cm}

\noindent {\it Rate Formulas under realistic (noisy) gate operations.} When realistic noisy operations are considered at the repeater nodes, the rate formulas in Table~\ref{tab:Rates_Ion_Requirements} get scaled by the distillable entanglement of the noisy end-to-end entangled state $\rho_{AB}$ across the line repeater network. The noisy entangled state is given by a Werner state of fidelity parameter $F=1-\frac{3}{2}Q(n)$, where $Q(n)$ is as given in Eqn.~\ref{eqn: end-to-end_Q}. A lower bound on the distillable entanglement is given by the reverse coherent information of the state $\rho_{AB}$, defined as $I_R(\rho_{AB}):=H(B)_\rho-H(AB)_\rho,$ where $H(B)_\rho=-\operatorname{Tr}(\rho_B\log_2\rho_B)$ is the von Neumann entropy of $\rho_B$. For Bell diagonal states, and hence for Werner states, $I_R$ can be easily computed, since $\rho_B$ is the maximally mixed state of entropy $H(B)_\rho=1$ and the entropy $H(AB)_\rho=-F\log_2F-(1-F)\log_2\frac{1-F}{3}$.

\vspace{0.5 cm}

\noindent {\it Ion Requirements:} For the case $T\geq\tau_0\geq \tau_g$, the ion requirements can be identified from the timing chart in Table~\ref{tab:Table_A}. 
The requirement on the number of communication ions is $2jM$, which is the value at which freed ions begin getting reused and the number of loaded $\textrm{Ba}^+$ ions saturates. In other words, $2jM$ $\textrm{Ba}^+$ ions are sufficient to support the optimal $(m,M)$ repeater protocol. 
The maximum number of memory ions required in this case is given by $2mM$. The actual number could be smaller, depending on the value of $m$ and its relation to $j,k$, which might allow for some freed memory ions to be reused. On the other hand, for the cases where $\tau_o>(k+j)\tau$, the communication ion requirement is $2(Mk+j)$, whereas the memory ion requirement is independent of $M,$ and given by $2m$. This is because the large $\tau_o$ allows one to wait for the heralding information and subsequently apply the swap gate only between the successfully heralded communication ion and the corresponding memory ions.

\begin{figure}
	\centering
	\includegraphics[width=0.45\textwidth]{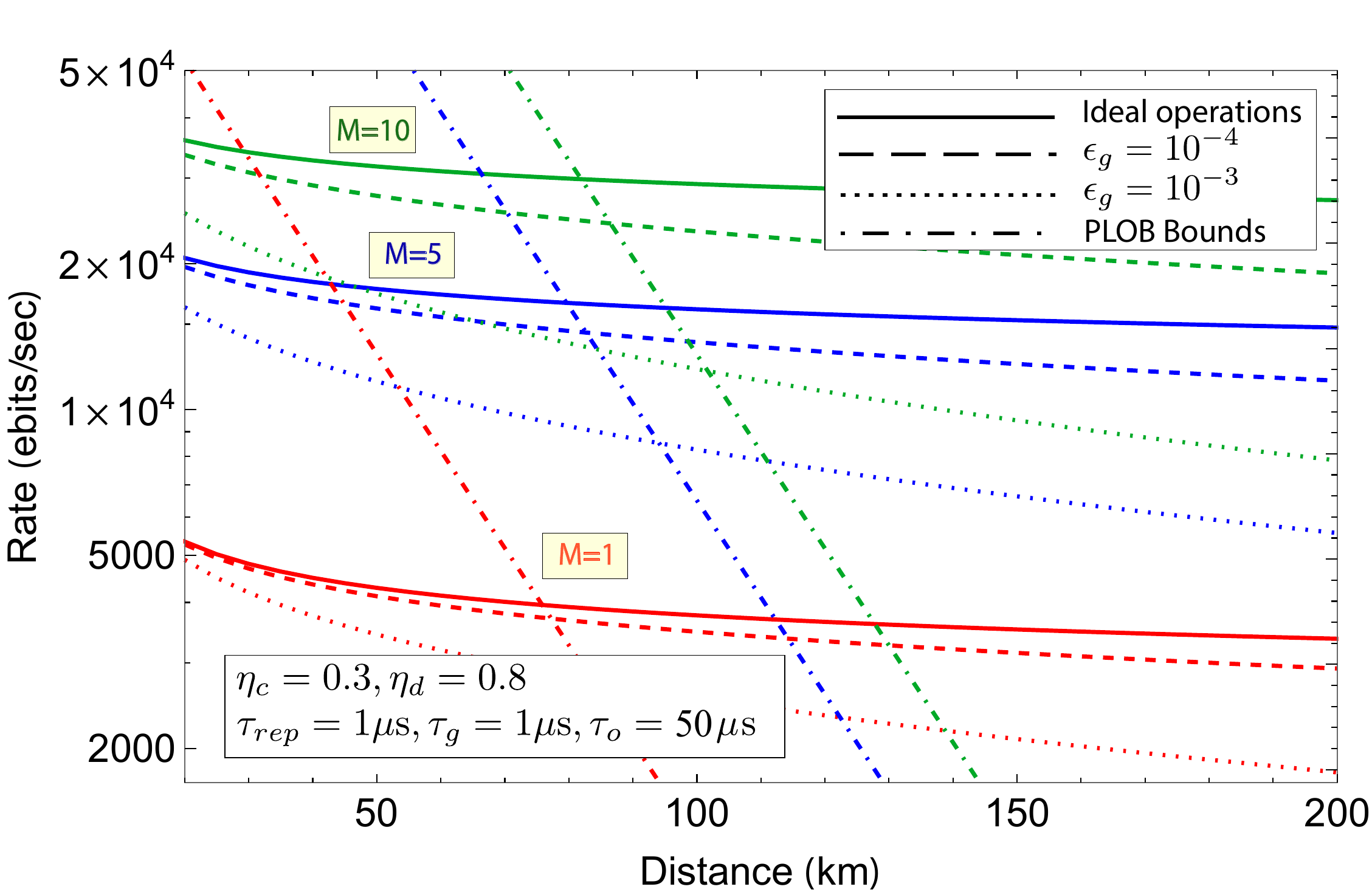}
	\caption{Entanglement distribution rate as a function of total distance optimized over the number of repeater nodes and the degree of time multiplexing, for different values of spatial multiplexing $M$, noise parameter $\epsilon_g$, and $\tau_g=\tau=1\mu$s. 
	These rates are compared against the direct transmission benchmark, namely the corresponding PLOB bounds (dotdashed lines) given by $-\frac{M}{\tau}\log_2(1-\eta).$}
	\label{fig:j1}
\end{figure}

\red{\subsection{Performance Evaluation: Numerical Results}}
\noindent {\it Unconstrained Repeaters:} Here we numerically analyze the performance of the repeater protocol assuming there are no constraints on the number of repeater nodes, or the number of ions per node. 
We begin with the rate-vs.-distance trade-off. 
To illustrate the results, we choose operating parameters of the repeater to be $\tau=1\mu$s, $\tau_g=1 \mu$s\ (i.e.,\ j=1), $\tau_o=50\mu$s, $\eta_c=0.3,\eta_d=0.8$, and the inter-repeater transmissions are assumed to be over optical fiber of attenuation $\alpha=0.2$ dB/km and refractive index $1.47$. 
The operational errors in gates and measurements are chosen as $\epsilon_g=1-F_0$, where $F_0$ is the Werner fidelity of the elementary link states. 
The value of $F_0$ is varied from $0$ (for ideal repeaters) to $10^{-4},10^{-3}$ (with noisy operations). 
Different values of spatial multiplexing $M=1,5,10$ are considered. 
The rates are numerically optimized over the time multiplexing block length $m$ and the number of repeaters $n$. 
The maximum of the optimal values of the different rate expressions corresponding to the different cases in Table~\ref{tab:Rates_Ion_Requirements} (which for the chosen set of parameters happens to correspond to Table~\ref{tab:Rates_Ion_Requirements} B, Case 2), is plotted in Fig.~\ref{fig:j1}. 
The rates are found to show sub-exponential decay with respect to distance, primarily owing to deterministic entanglement swapping and additionally due to time multiplexing. 
The rates are higher for higher $M$, but the advantage over the corresponding PLOB bounds calculated as per Eqn.~\ref{PLOB_ebits/sec} also occurs at commensurately longer distances. 
In the presence of operational errors in the repeaters, the degradation of the rates with distance is more pronounced with increasing values of the noise parameter. 
Nevertheless, the rate-distance trade-off still beats the PLOB bound.

\begin{figure}
	\centering
	\includegraphics[width=0.45\textwidth]{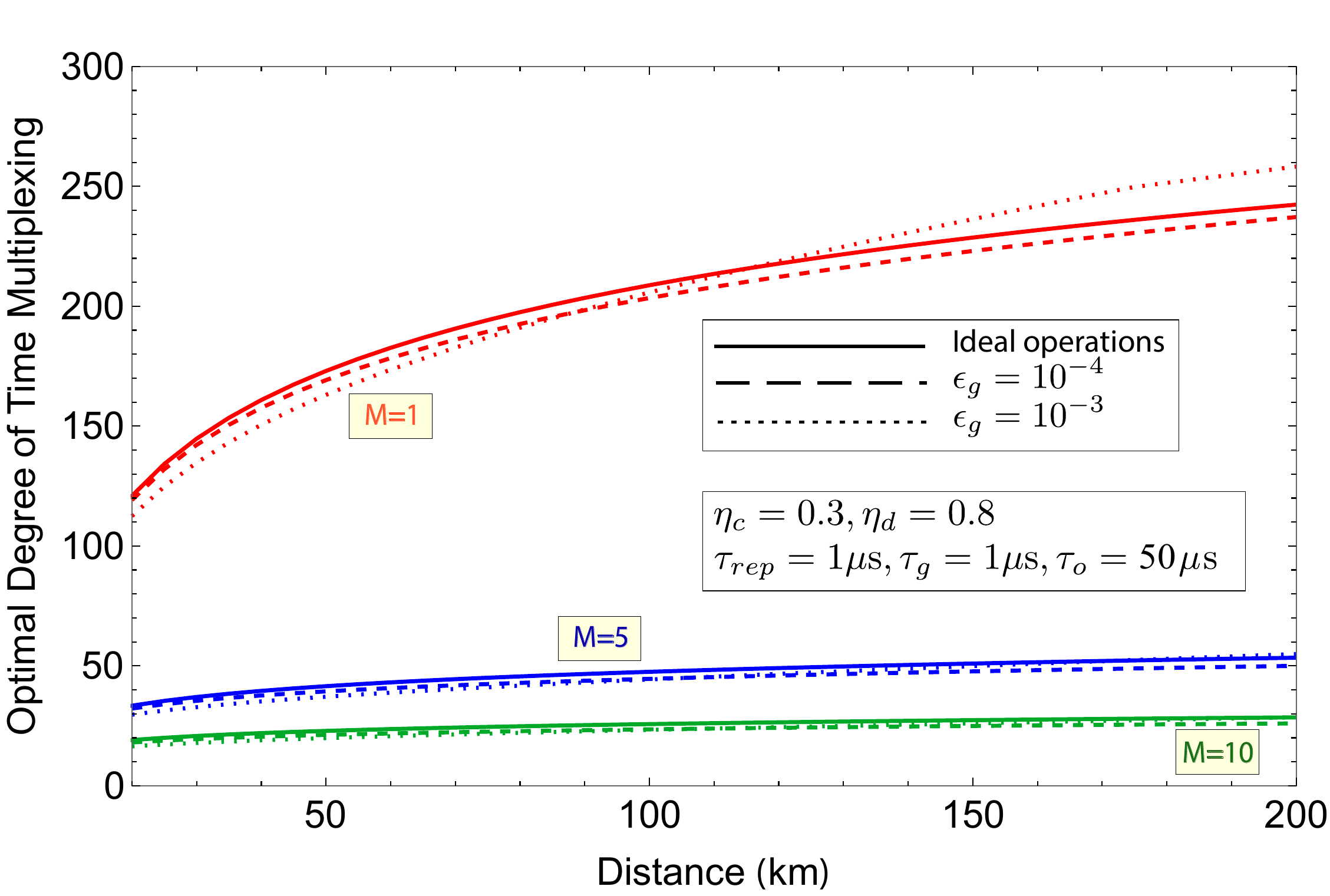}
	\caption{Optimal degree of time multiplexing as a function of total distance for different values of spatial multiplexing $M$, noise parameter $\epsilon_g$, $\tau_g=\tau=1\mu$s, and optimal number of repeaters.}
	\label{fig:j_m_opt}
\end{figure}

\begin{figure}
	\centering
	\includegraphics[width=0.45\textwidth]{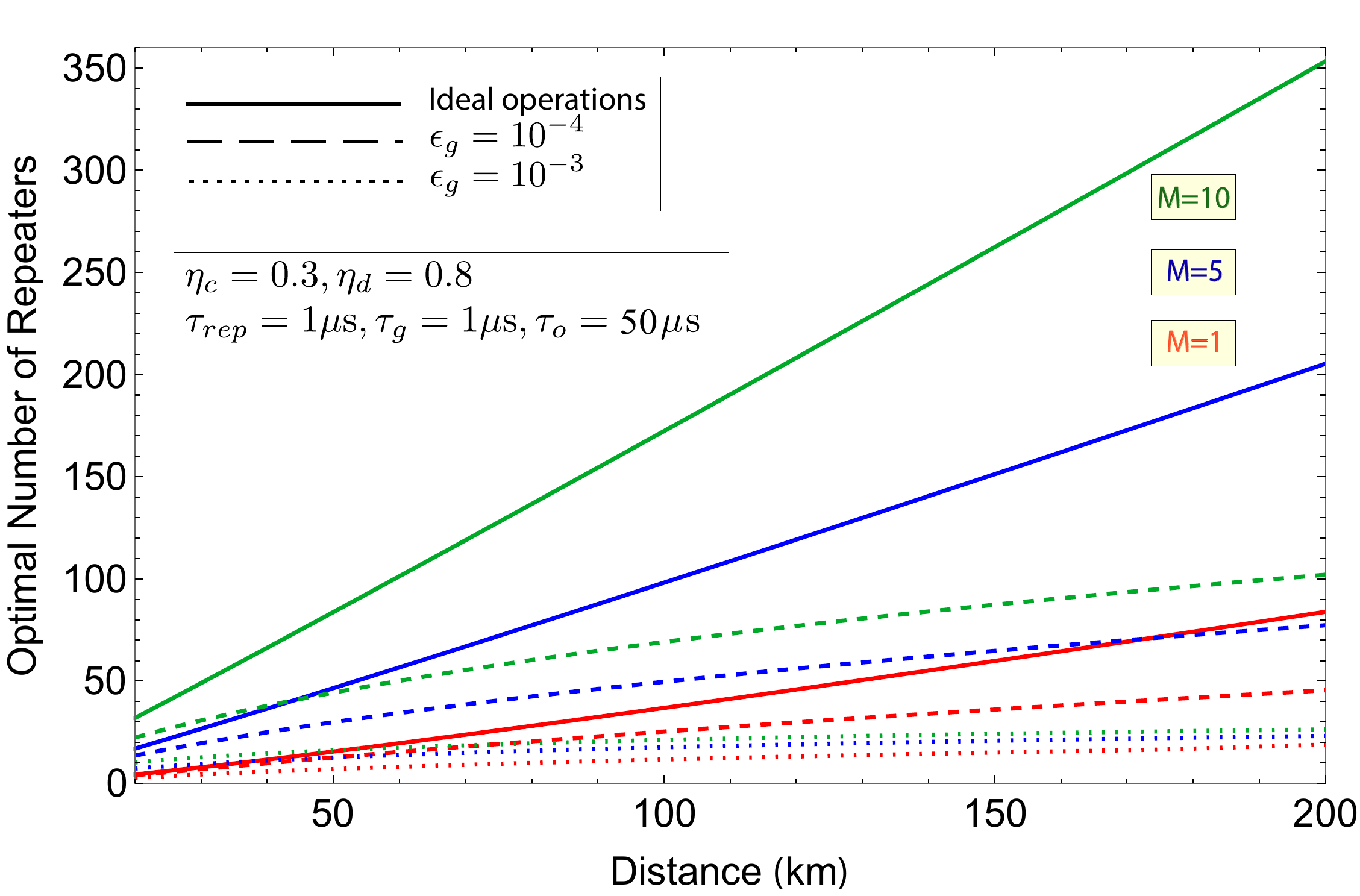}
	\caption{Optimal number of repeater nodes as a function of total distance for different values of spatial multiplexing $M$, noise parameter $\epsilon_g$, $\tau_g=\tau=1\mu$s and optimal time multiplexing.}
	\label{fig:j_n_opt}
\end{figure}

The optimal time multiplexing block-length and the optimal number of repeaters for different degrees of spatial multiplexing $M$ are plotted as functions of the total distance in Figs.~\ref{fig:j_m_opt} and \ref{fig:j_n_opt}, respectively. 
Notice that the optimal value of $m$ increases with distance, saturating at large distances. 
The optimal value of $m$ in relation to noise in the gates and measurements is found to behave non-monotonically, first decreasing then increasing with noise at large distances. 
However, most importantly, the values of $m$ are higher for the lower value of $M$. 
In fact, at any given distance $L$ and noise parameter, when $M$ is varied, the optimal $m$ ( call it $m_\textrm{opt}$) satisfies the same mode multiplexing product $\mathbf{m}=m \times M$. 
For a total distance of $150$ km and noise parameter $\epsilon_g\leq 10^{-4}$, the optimal product is found to be $\mathbf{m}_{\textrm{opt}}\approx 220$. 
\red{The reason we find an optimal value for the product is that the rate (in ebits/sec) per spatial mode, i.e., the rate in Eqn. (\ref{mMRate}) divided by $M$, has a direct dependence on the product of the two multiplexing parameters, i.e., $m \times M$. 
(The value $220$ itself is a function of the choice of system parameters such as the communication ion lifetime, gate times, optical fiber attenuation, and coupling and detection efficiencies.)} 
As a result, for $M=1,5,10$, we have $m_\textrm{opt}=220,44,22$, respectively. 
The optimal number of repeaters is seen to grow with the total distance at a rate proportional to $M$. 
It is found to naturally slow down with increasing gate and measurement noise, as more QR nodes would add more operational noise to the shared ebits.
For example, for $M=10$ and a total distance of $L=150$ km, the optimal number of repeaters for noise parameter values $\epsilon_g=10^{-4},10^{-3}$, are found to be $87$ and $25$, respectively, which amount to inter-repeater spacing values of $L_0\approx 1.7$ km and $6$ km, respectively.

\begin{figure}
	\centering
	\includegraphics[width=0.45\textwidth]{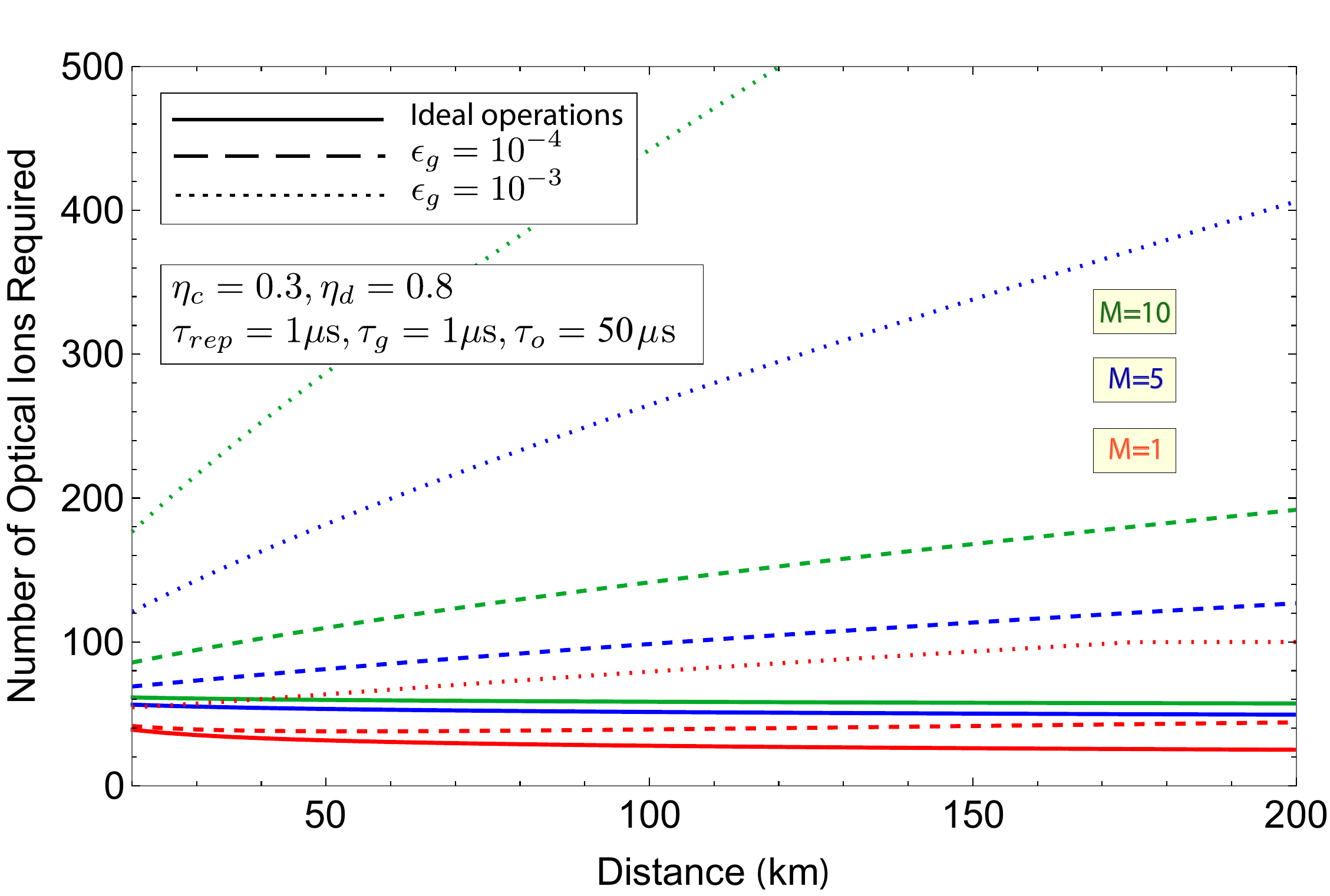}
	\caption{Required number of communication ions as a function of total distance for different values of spatial multiplexing $M$, noise parameter $\epsilon_g$, $\tau_g=\tau=1\mu$s and optimal number of repeaters and time multiplexing.}
	\label{fig: N_o_req}
\end{figure}
\begin{figure}
	\centering
	\includegraphics[width=0.45\textwidth]{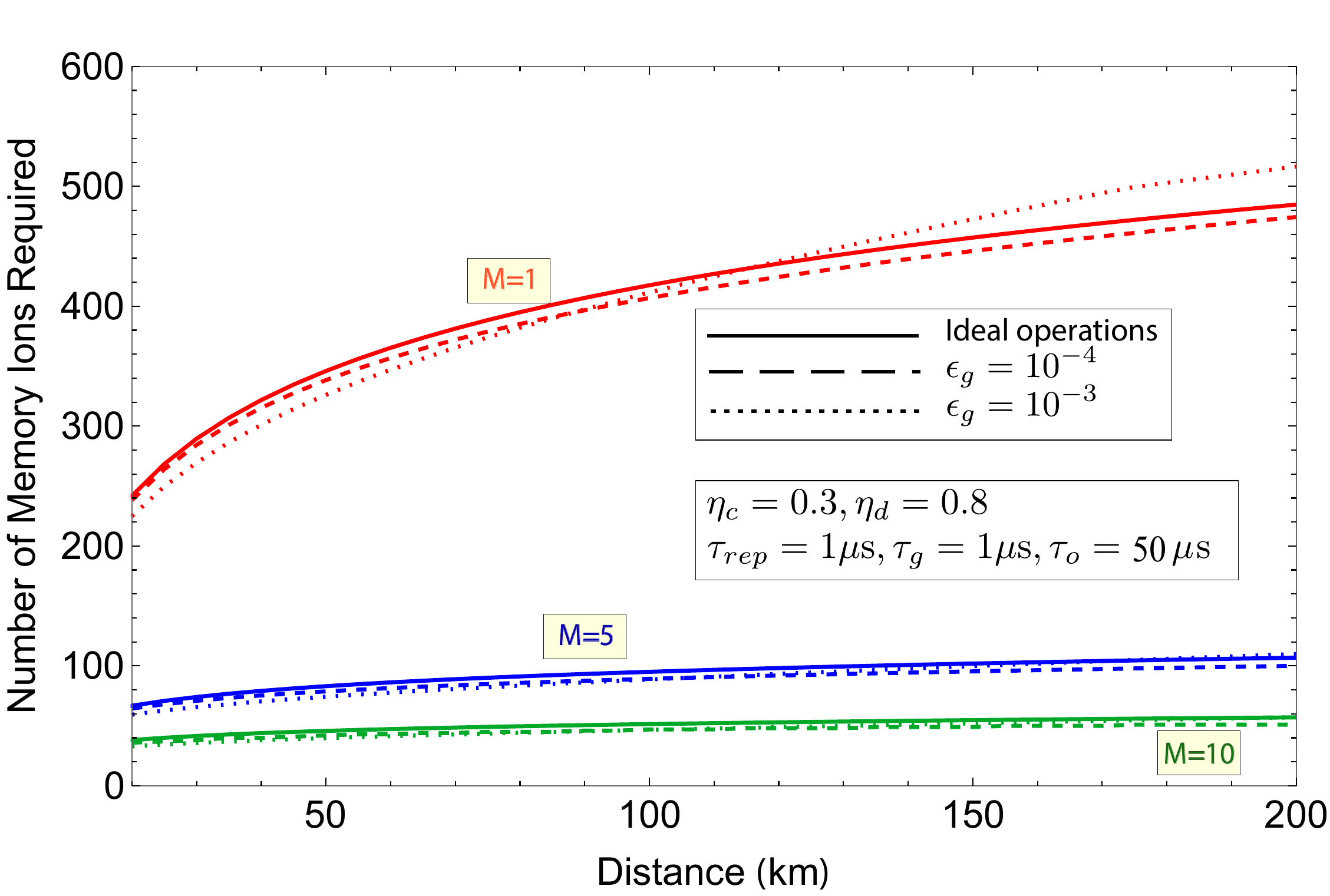}
	\caption{Required number of memory ions as a function of total distance for different values of spatial multiplexing $M$, noise parameter $\epsilon_g$, $\tau_g=\tau=1\mu$s and optimal number of repeaters and time multiplexing.}
	\label{fig: N_m_req}
\end{figure}
The number of communication and memory ions per repeater node $(N_o,N_m)$ required to support the optimal rates under the proposed mode-multiplexing protocol are shown in Figs.~\ref{fig: N_o_req} and \ref{fig: N_m_req}, respectively. 
The value of $N_o$ decreases sub-exponentially with distance so long as the gate and measurement noise $\epsilon_g$ is small. 
In such a scenario, the required numbers are smallest for protocol with smaller $M$ and increases with $M$. 
The value of $N_o$ tends to increase with distance for $\epsilon_g$ above a threshold. 
This is because large $\epsilon_g$ drives down the optimal number of repeaters and consequently drives up the inter-repeater spacing. 
With increasing inter-repeater spacing, the protocol in Table~\ref{tab:Rates_Ion_Requirements} B, Case 2 warrants higher number of communication ions given by $2(MT+\tau_g)/\tau$.
On the other hand, the required number memory ions always increases with distance, since the optimal time multiplexing blocklength $m$ increases, too, and the required number of memory ions is proportional to $m$. 
It is higher for lower values of $M$ (or in other words for higher values of $m$).

\begin{figure}
	\centering
	\includegraphics[width=0.45\textwidth]{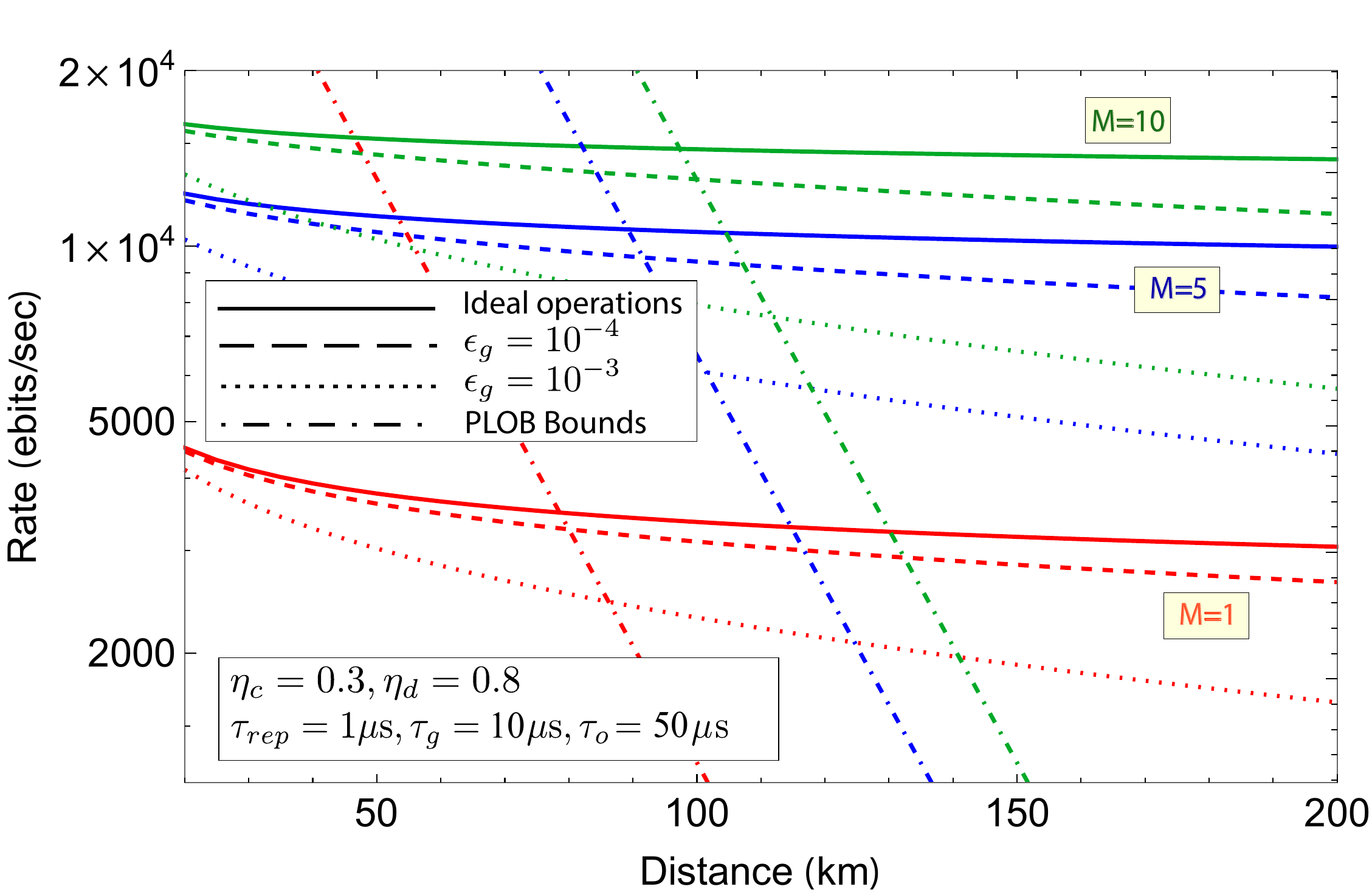}
	\caption{Entanglement distribution rate as a function of total distance optimized over the number of repeater nodes and the degree of time multiplexing, for different values of spatial multiplexing $M$, with realistic noisy gate operations of infidelity $\epsilon_g=10^{-4}, \tau=1\mu$s and $\tau_g=10\tau=10\mu$s. These rates are compared against the direct transmission benchmark, namely the corresponding PLOB bounds (dot dashed lines) given by $-\frac{M}{\tau}\log_2(1-\eta).$}
	\label{fig:j1vs10}
\end{figure}
\vspace{0.5 cm}


We note that the rates attained in Fig.~\ref{fig:j1} based on the protocol in Sec.~\ref{sec: protocols} are higher than those reported in ref.~\cite[Fig.~7(b)]{Santra2019-di}. 
For instance, at a total distance of $150$ km, noise parameter $\epsilon_g=10^{-4}$, $M=10$ spatial multiplexing, and all other parameters being identical, our protocol attains $20000$ ebits/sec, whereas the protocol in ref.~\cite{Santra2019-di} achieves 700 ebits/s. 
The rate enhancement does come at the cost of higher ion number requirements. 
For the said parameter values, the required number of communication ions and memory ions per repeater for our protocol are $N_o=170$, and $N_m=55$, respectively, whereas the protocol in ref.~\cite{Santra2019-di} required only $N_o=10, N_m=2$.

To conclude this section, in Fig.~\ref{fig:j1vs10}, we plot the entanglement distribution rates when $\tau=1\mu$s, and $\tau_g=10\mu$s, i.e., for $j=10$. In other words, this refers to a scenario where the gate operations are an order of magnitude slower compared to the clock-cycle duration for ion-photon entanglement generation, which is still retained at $1\mu$s. 
The end-to-end entanglement distribution rates are seen to decrease only marginally (compared to the case $\tau_g=\tau=1\mu$s). 
This is made possible by an increased requirement on the number of communication ions in the DSTI modules. 
For example, for $M=10, \epsilon_g=10^{-4}$, while the required number of communication ions in the case of $\tau_g=\tau=1\mu$s was 170, it is 220 for the case $\tau_g=10\tau=10\mu$s. 
Further, the repeater operation in the latter scenario warrants traps with distinct sectors of communication ions ($j=\tau_g/\tau=10$ number of sectors) that can be excited successively in a cyclical manner.

\vspace{0.5 cm}
\noindent {\it Constrained Repeaters:} \red{Here we numerically analyze how restrictions on the number of repeater nodes and on the amount of ion resources at the repeaters affect the entanglement distribution rates supported by the proposed protocol.} 

\begin{figure}
	\centering
	\includegraphics[width=0.45\textwidth]{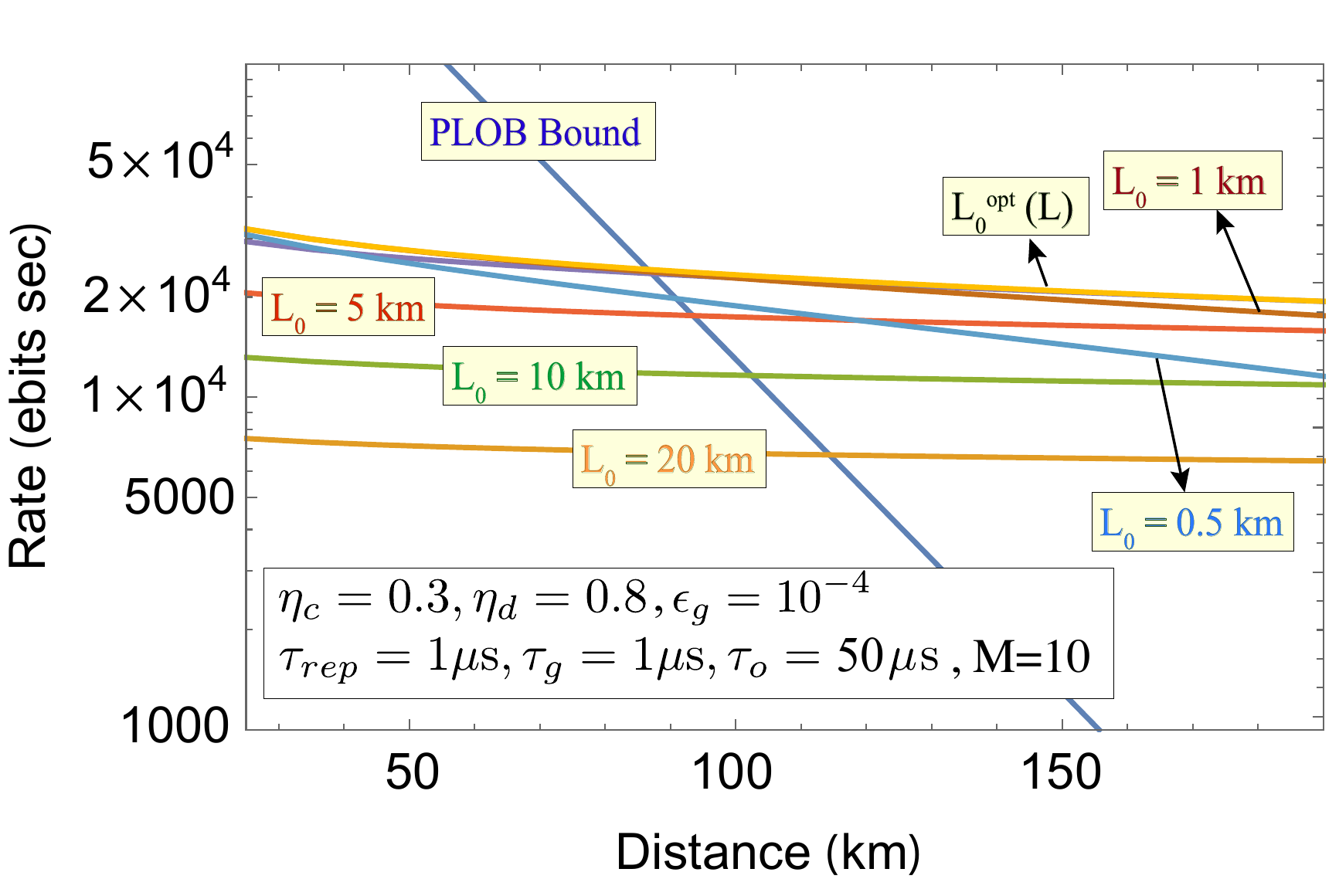}
	\caption{\red{Entanglement distribution rate as a function of total distance optimized over the degree of time multiplexing, for spatial multiplexing degree $M=10$, noise parameter $\epsilon_g=10^{-4}$, $\tau_g=\tau=1\mu$s, and different values of inter-repeater spacing $L_0$. 
	These rates are compared against the direct transmission benchmark, namely the corresponding PLOB bound given by $-\frac{M}{\tau}\log_2(1-\eta).$}}
	\label{fig: fixed_l0_rates}
\end{figure}

\red{To understand how the entanglement distribution rates deteriorate upon moving away from the optimal number of repeaters, the rates for different but fixed values of inter-repeater spacing $L_0$ are plotted in Fig.~\ref{fig: fixed_l0_rates} for repeater operations with $M=10$ and $\epsilon_g=10^{-4}$. 
For $L_0$ larger than the distance-dependent $L_0^{\textrm{opt}}$, the rates are seen to be significantly smaller. 
For example, with $L_0=20$ km, the rate at a total distance of $150$ km drops to $\approx 25\%$ of its value corresponding to the optimal value of $L_0\approx 1.7$ km.
However, the rate-distance scaling remains unchanged in the limit of large distances, which implies it is still possible to operate at rates higher than direct transmission rates even with fewer and farther-spaced repeaters at large distances.}

\begin{figure}
	\centering
	\includegraphics[width=0.45\textwidth]{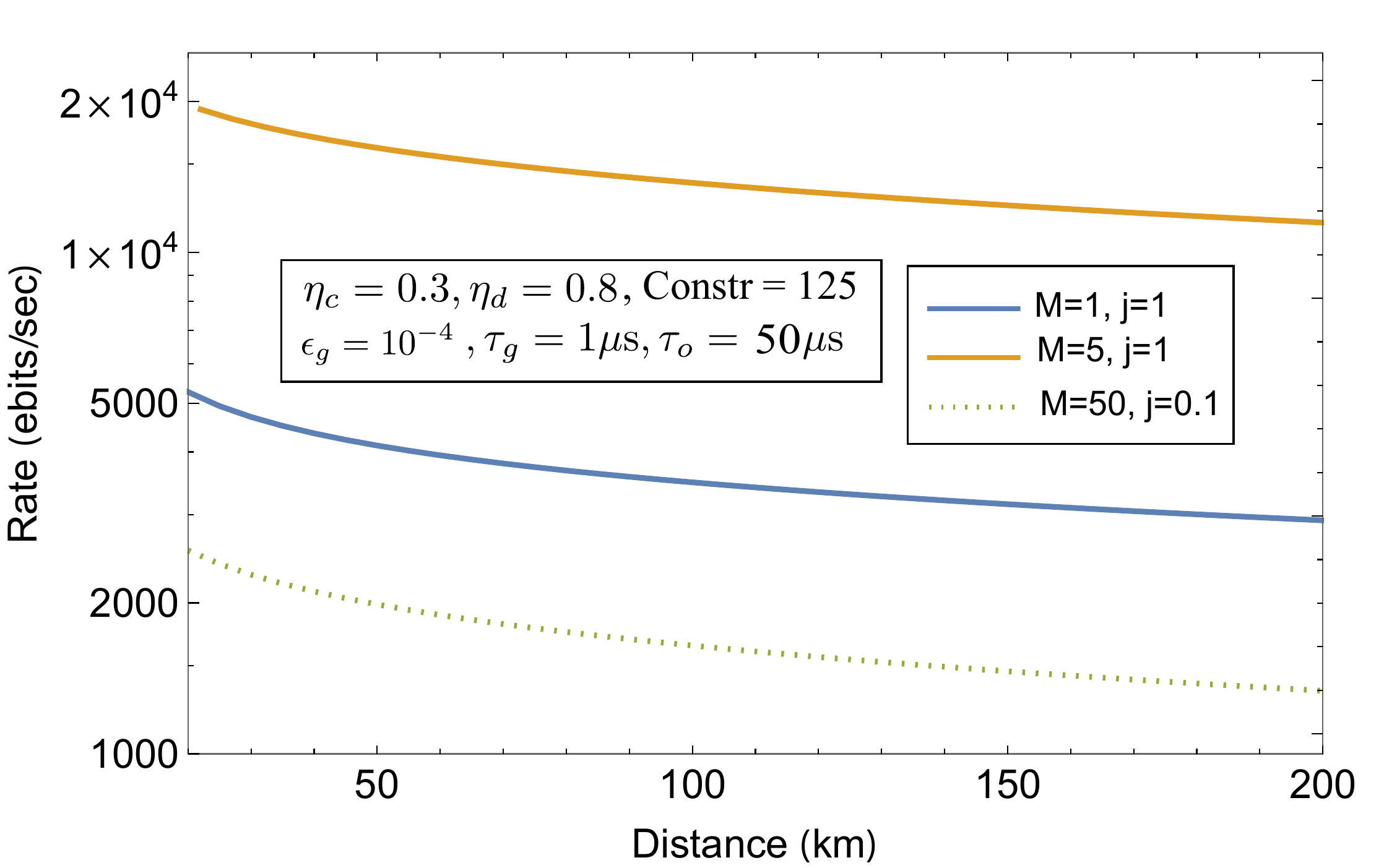}
	\caption{Entanglement distribution rate as a function of total distance under constrained resource availability, $N^{\max}_o=125$. The values of $M$ are specified with realistic noisy gate operations of infidelity $\epsilon_g=10^{-4}$ and $\tau_g=1\mu$s. The operational clock cycle duration $\tau$ is different for the solid $(\tau=1\mu s)$ and dotted $(\tau=10\mu s)$ lines.}
	\label{fig:consper1}
\end{figure}

With regard to the number of communication ions, packing too many of these ions in a DSTI module can cause issues with ion-photon entanglement due to resonant re-absorption of emitted photons by neighboring communication ions. 
Hence, the effect of restrictions on the number of communication ions is an important consideration.
Given a limit on the maximum number of communication ions $N^{\max}_o$, a fixed gate operation time $\tau_g$ and the minimum allowed clock cycle duration $\tau^{\min}$, the rates can be optimized over the number of repeaters and time multiplexing parameter for different combinations of spatial multiplexing $M$ and $j$ where $j=\tau_g/\tau$ can be varied by choosing different values for the clock cycle duration $\tau$. 
For $N^{\max}_o=125$, $\tau_g=\tau^{\min}=1\mu s$ and noise parameter $\epsilon_g=10^{-4}$, 
Fig.~\ref{fig:consper1} shows the plots of the repeater performance for a few different allowed $M,j$ combinations, when the number of optical ions is constrained to 125. 
The solid lines show the repeater performance when $j=1$, which appears to allow a maximum $M$ of 5. 
By changing $\tau$ to $10 \mu s$, i.e., by reducing $j=0.1$, the maximum allowable $M$ increases to 50. 
The performance under this modified regime of timing parameters is shown by the dashed line, which is clearly sub-optimal. 
Thus, a higher degree of spatial multiplexing at the expense of slower clock rate $1/\tau$ does not appear to yield higher rates. 
Instead, the optimal strategy appears to be to pick the highest possible $j$ value and then optimally choosing $M$ such that the constraint $N^{\max}_o$ is still satisfied.

\begin{figure}
	\centering
	\includegraphics[width=0.45\textwidth]{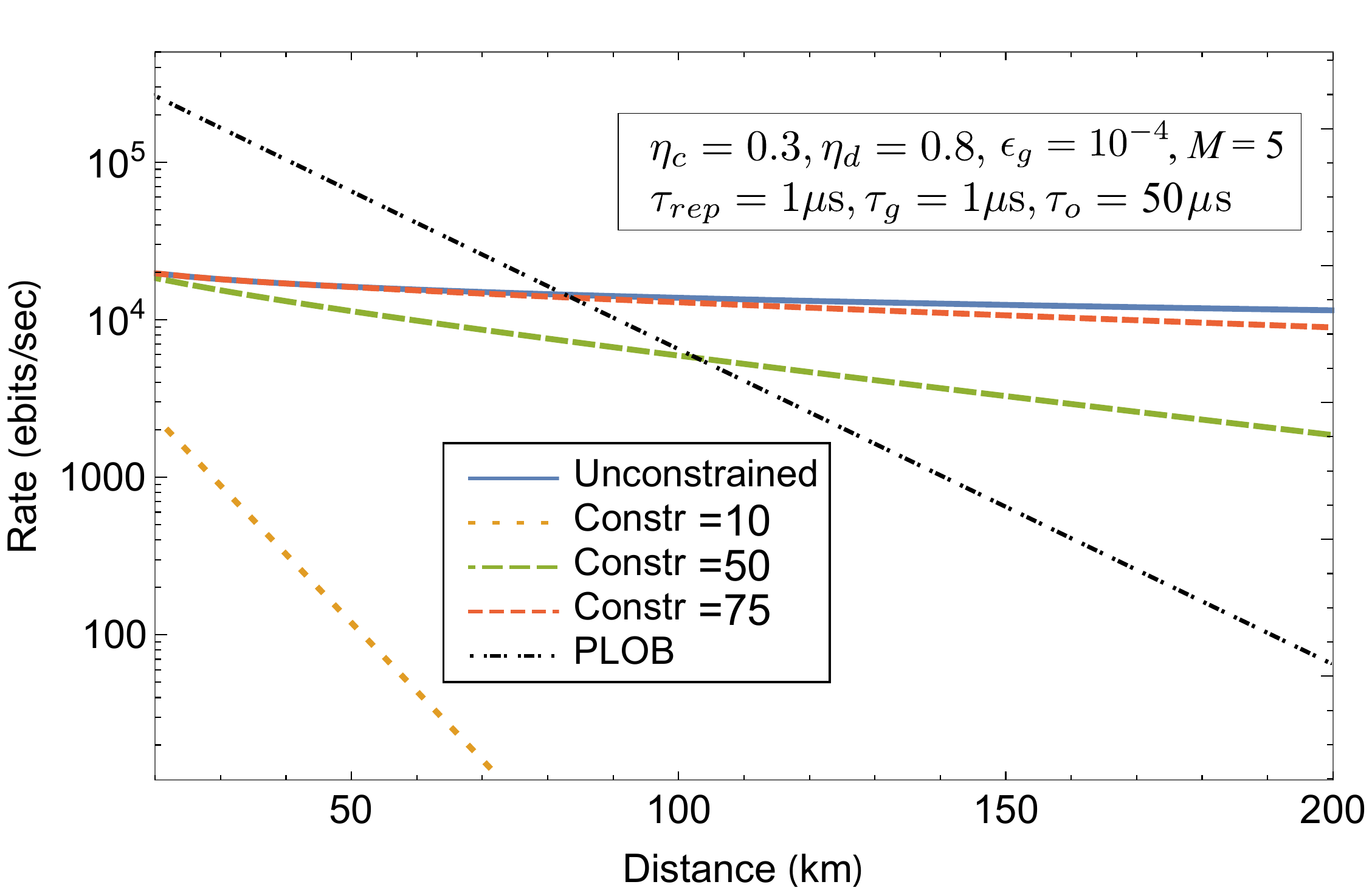}
	\caption{Entanglement distribution rate as a function of total distance under constrained resource availability, $N^{\max}_m$. For spatial multiplexing $M=5$, realistic noisy gate operations of infidelity $\epsilon_g=10^{-4}$ and $\tau_g=\tau=1\mu$s, the rates corresponding to different values of $N^{\max}_m$ are plotted.}
	\label{fig:consper2}
\end{figure}

Finally, with regard to the number of memory ions, given a limit on the maximum number of these ions, the set of allowed values for the time multiplexing parameter becomes restricted. 
The reader may refer to the steady state value of ion occupancy values in Tables~\ref{tab:Table_A}-\ref{tab:Table_C} to understand the dependence of various constraints on the protocol and network parameters. 
As a result the optimal rate drops faster with distance for smaller values of $N^{\max}_m$ as illustrated in Fig.~\ref{fig:consper2}. 
When $N^{\max}_m$ falls below a threshold, the rate-distance scaling no longer beats the direct transmission benchmark.


\section{Conclusion}
\label{sec: discussion}

To conclude, we presented a general architecture for a repeater node based on DSTI modules, and discussed a repeater protocol based on spatial and time multiplexing. For DSTI modules with $^{138}$Ba$^+$ as communication ions and $^{171}$Yb$^+$ as memory ions, assuming reasonable values for operating parameters (operation errors under $\epsilon_g<10^{-4}$, gate time $\tau_g=1\mu$s, clock cycle duration $\tau=1\mu$s, communication ion lifetime $\tau_o=50\mu$s, coupling efficiency $\eta_c=0.3$ and detection efficiency $\eta_d=0.8$), the proposed repeater protocols based on spatial and time multiplexing can attain entanglement distribution rate $\sim 20000$ ebits/s at a distance of 150 km, with repeaters placed at $\approx 1.7$ km spacing, and each containing about $170$ and $55$ $^{138}$Ba$^+$ and $^{171}$Yb$^+$ ions, respectively. 
This constitutes a nearly $30\times$ improvement over the rate reported in the earlier work of \cite{Santra2019-di} for the same set of operating parameters, but requires larger number of ions at the repeater nodes. 
The larger ion number requirements can potentially be met by bootstrapping several DSTI modules at the repeater nodes. 
However, the modular interactions even within a repeater node would then require probabilistic optically-mediated entanglement swapping operations, which can cause a degradation of the entanglement distribution rates. 
This calls for the design of more advanced protocol that can assuage this degradation and optimally leverage multiple DSTI modules at repeater nodes. 
In this regard, allowing for arbitrary optically-heralded intra-node ion-ion logic across traps along with universal-capable logic in the traps, we will explore protocols that incorporate block entanglement distillation codes~\cite{BDSW96} as part of future work. These protocols will take advantage of multiple successful, elementary entanglement generation attempts, while alleviating the drawbacks stemming from the constraint of having to measure all the ions in a trap simultaneously and the probabilistic, optically heralded intra-node logic.

\begin{acknowledgments}
This work was supported by the National Science Foundation Convergence Accelerator Program, award number 2040695. 
NML additionally acknowledges support from the Army Research Office, award number W911NF1910296. 
SG and KPS additionally acknowledge the NSF Center for Quantum Networks (CQN), grant number EEC-1941583. 
KPS thanks Liang Jiang for helpful discussions.
\end{acknowledgments}

\nocite{*} 
\bibliography{bibliography}
\bibliographystyle{unsrt}

\onecolumngrid

\appendix
\section{Timing Analysis for Resource Count Calculation}
\label{appendix:timing}
\begin{figure}[h]
	\centering
	\includegraphics[width=0.6\textwidth]{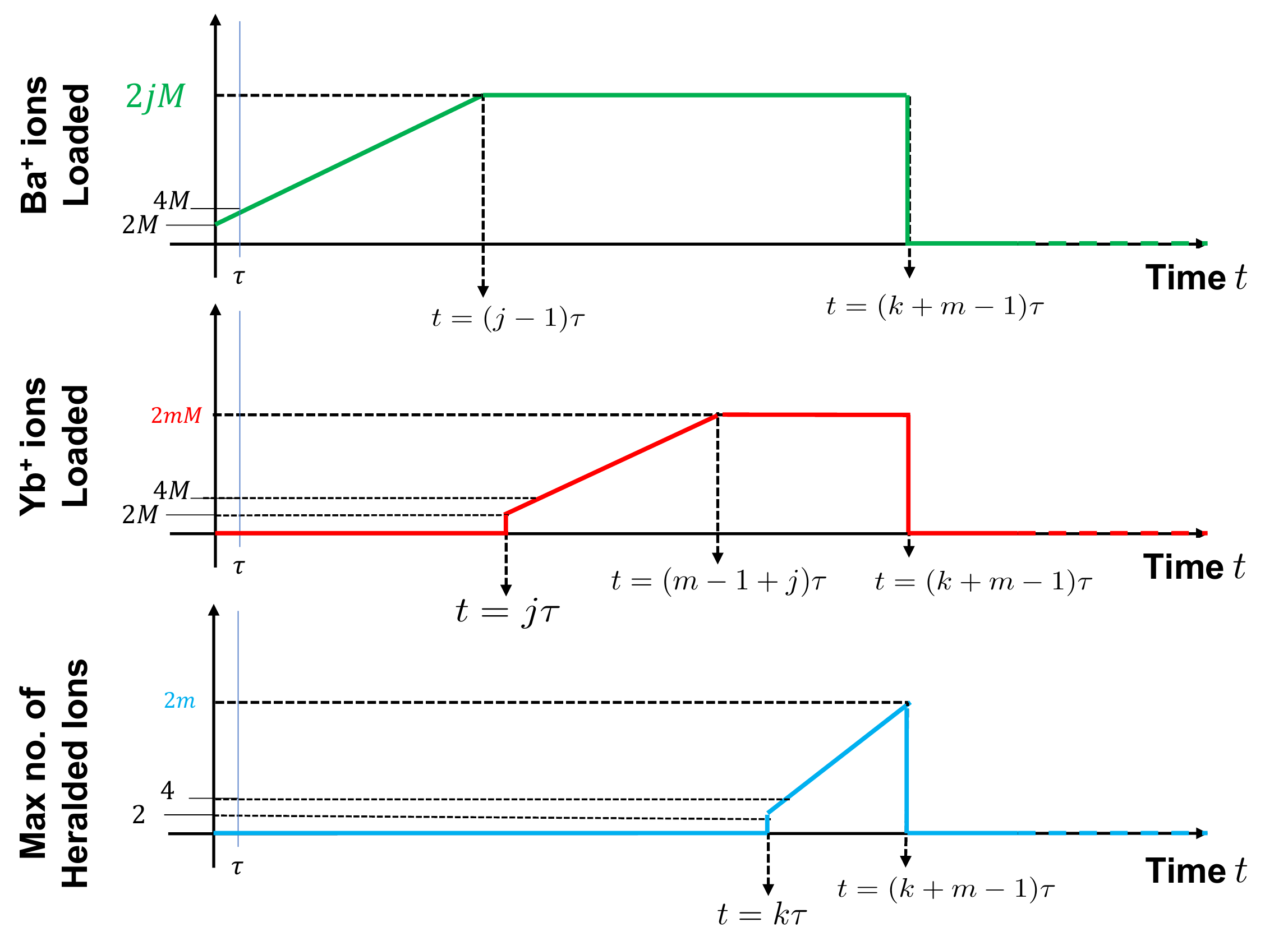}
	\caption{Timing diagram for Table~\ref{tab:timing_parameters} Type A when $k-j+1\geq m$. 
	}
	\label{fig:timing_typeAc1}
\end{figure}

\begin{figure}[h]
	\centering
	\includegraphics[width=0.6\textwidth]{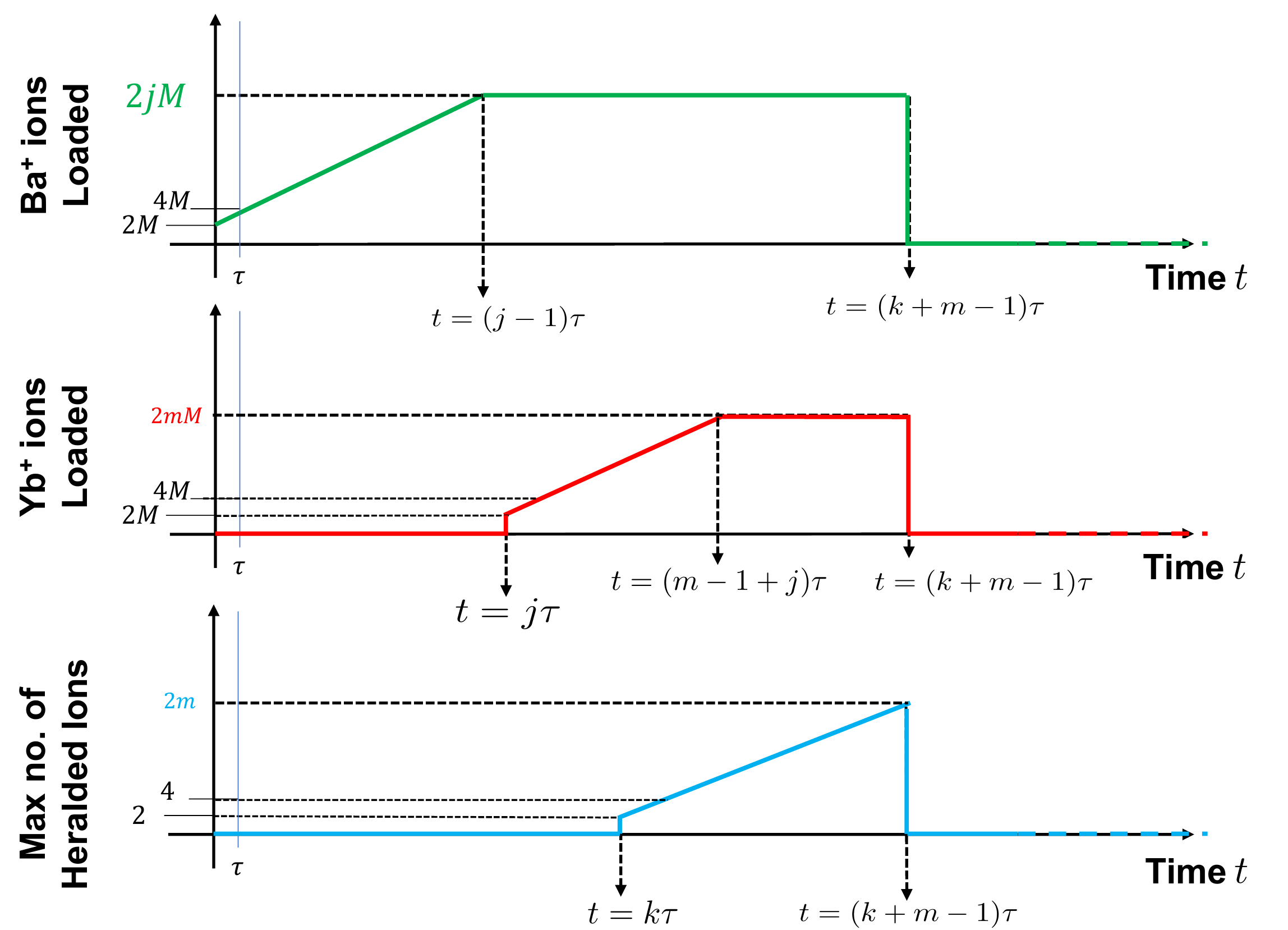}
	\caption{Timing diagram for Table~\ref{tab:timing_parameters} Type A when $k-j+1<m$.}
	\label{fig:timing_typeAc2}
\end{figure}

\begin{figure}[h]
	\centering
	\includegraphics[width=0.6\textwidth]{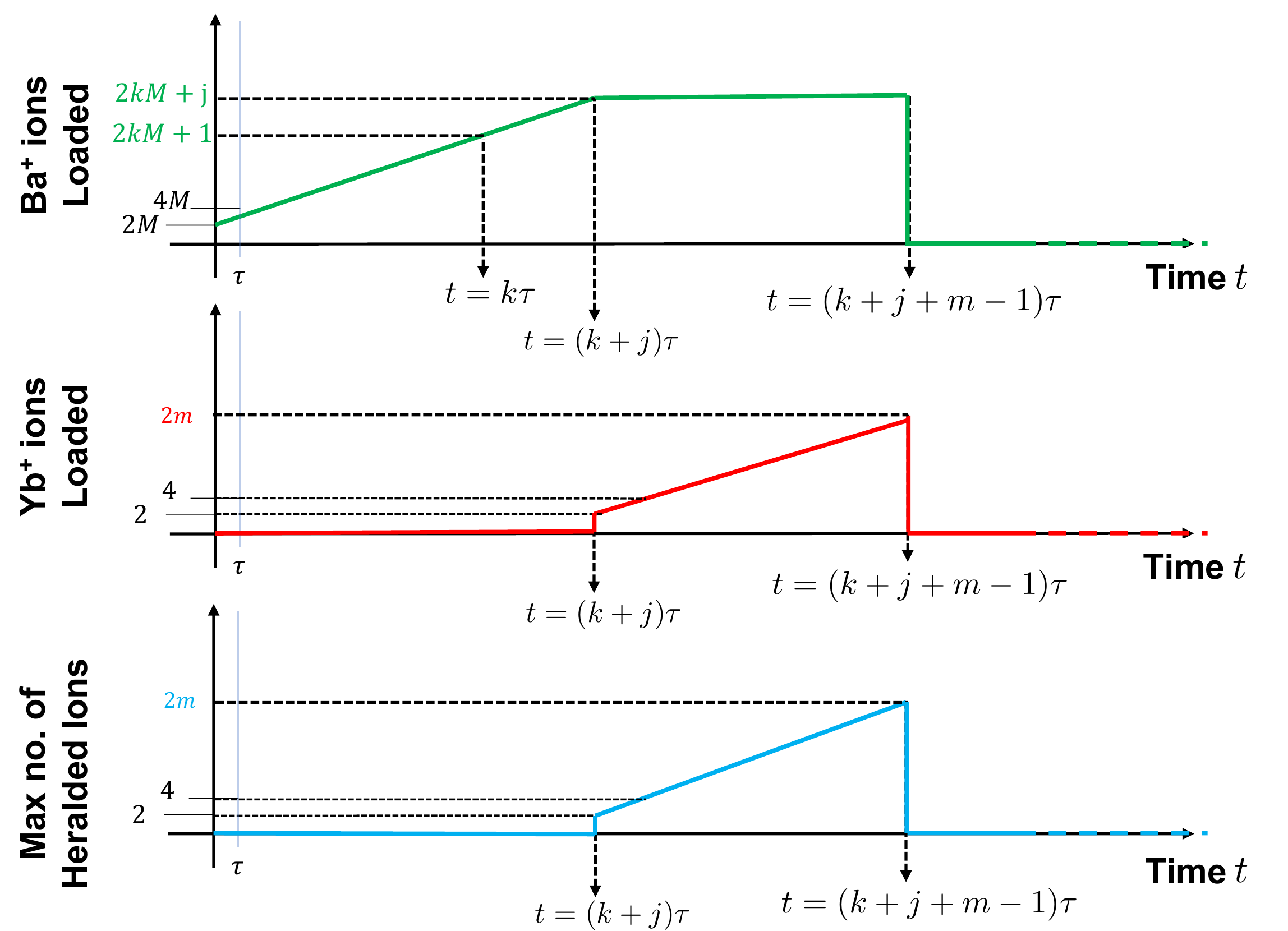}
	\caption{Timing diagram for Table~\ref{tab:timing_parameters} Type B.}
	\label{fig:timing_typeB}
\end{figure}

\begin{figure}[h]
	\centering
	\includegraphics[width=0.6\textwidth]{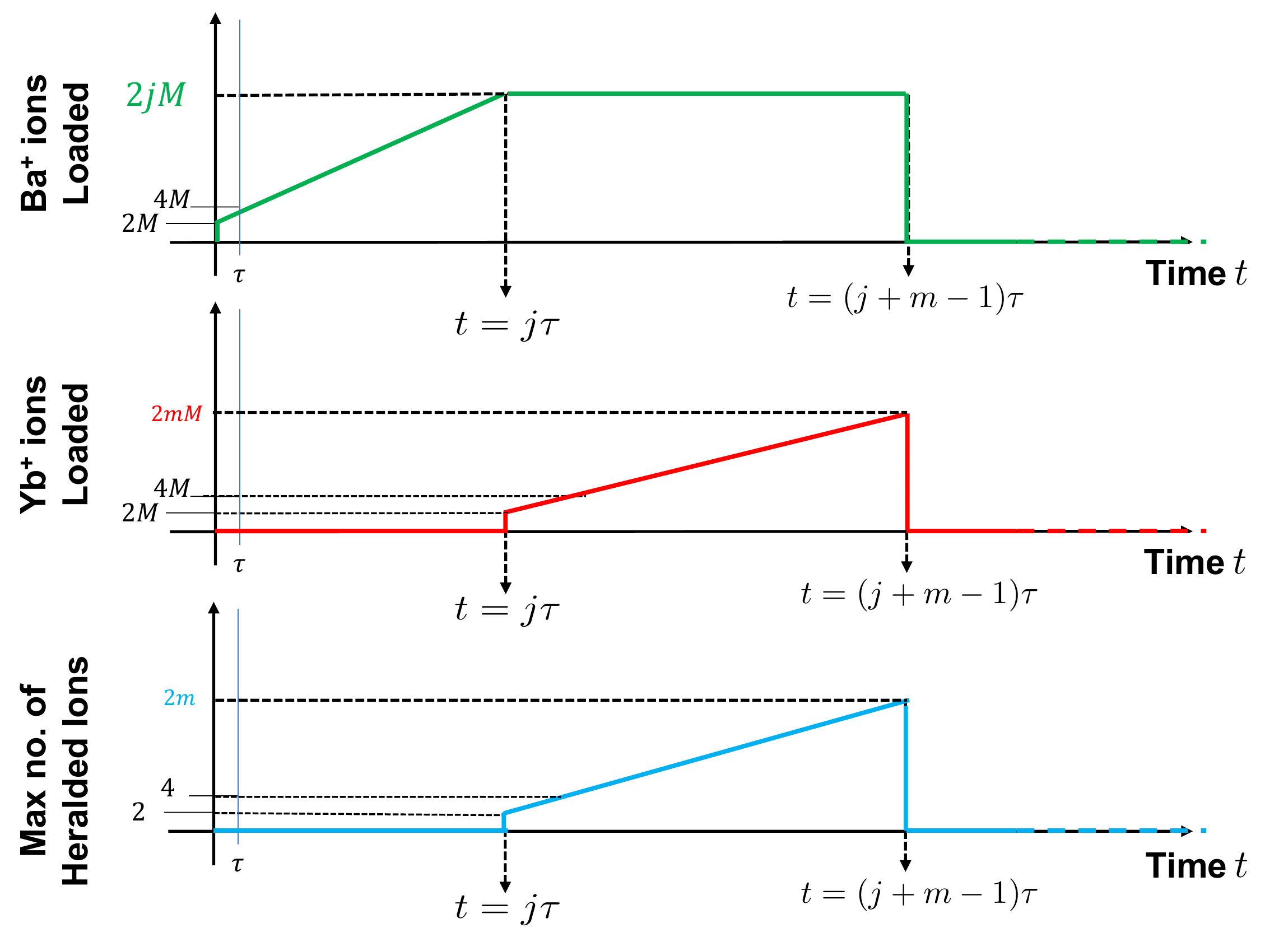}
	\caption{Timing diagram for Table~\ref{tab:timing_parameters} Type C.}
	\label{fig:timing_typeC}
\end{figure}

There are multiple operating regimes for realistic operations with noisy and non-instantaneous quantum gates. The reader may refer to Table~\ref{tab:Rates_Ion_Requirements} for a summary of the various protocol types. We have examined the timing analysis of each protocol type in depth in Tables~\ref{tab:Table_A}-\ref{tab:Table_C}. The corresponding timing diagrams are shown in Figs.~\ref{fig:timing_typeAc1}-\ref{fig:timing_typeC}

\begin{table}[ht]
	\centering
	\begin{tabular}{c c c c c c c}
		\toprule
		Case 1 & \multicolumn{3}{c}{$Ba^+$ Occupancy} & \multicolumn{2}{c}{$Yb^+$ Occupancy} & \multirow{2}{3cm}{\centering Max. number of heralded ions}
		\\\cmidrule(lr){1-1}\cmidrule(lr){2-4}\cmidrule(lr){5-6}
		Time & Initialized  & Freed  & Loaded & Loaded  & Freed  &\\
		\midrule\\
		0 & $2M $ & - & $2M $ & - & - & - \\
		$\tau$ & $4M $ & - & $4M $ & - & - & - \\
		\vdots & \vdots & \vdots & \vdots & \vdots & \vdots & \vdots\\
		$ (j-1)\tau $ & $2jM $ & - & $\mathbf{2jM} $ & - & - & - \\
		$j\tau$ & $2(j+1)M $ & $ 2M $ & $2jM $& $ 2M $ & - & - \\
		\vdots & \vdots & \vdots & \vdots & \vdots & \vdots & \vdots\\
		$(m-1+j)\tau$ & $2(m+j)M $ & $ 2mM $ & $2jM $& $ \mathbf{2mM} $ & - & - \\
		\vdots & \vdots & \vdots & \vdots & \vdots & \vdots & \vdots\\
		$k\tau$ & $2(k+1)M $ & $ 2(k-j+1)M $ & $2jM $& $ 2(k-j+1)M $ &  $ 2(M-1) $& $ 2 $ \\
		\vdots & \vdots & \vdots & \vdots & \vdots & \vdots & \vdots\\
		$\mathbf{(k+m-1)\tau}$ & $2(k+m)M $ & $ 2(k+m-j)M $ & $2jM$ & $ 2(k+m-j)M $ &  $ 2m(M-1) $& $ \mathbf{2m} $ \\
		\bottomrule
	\end{tabular}
	
	\begin{tabular}{c c c c c c c}
		\toprule
		Case 2 & \multicolumn{3}{c}{$Ba^+$ Occupancy} & \multicolumn{2}{c}{$Yb^+$ Occupancy} & \multirow{2}{3cm}{\centering Max. number of heralded ions}
		\\\cmidrule(lr){1-1}\cmidrule(lr){2-4}\cmidrule(lr){5-6}
		Time & Initialized & Freed  & Loaded & Loaded  & Freed  &\\
		\midrule\\
		0 & $2M $ & - & $2M $ & - & - & - \\
		$\tau$ & $4M $ & - & $4M $ & - & - & - \\
		\vdots & \vdots & \vdots & \vdots & \vdots & \vdots & \vdots\\
		$ (j-1)\tau $ & $2jM $ & - & $\mathbf{2jM} $ & - & - & - \\
		$j\tau$ & $2(j+1)M $ & $ 2M $ & $2jM $& $ 2M $ & - & - \\
		\vdots & \vdots & \vdots & \vdots & \vdots & \vdots & \vdots\\
		$k\tau$ & $2(k+1)M $ & $ 2(k-j+1)M $ & $2jM $& $ 2(k-j+1)M $ &  $ 2(M-1) $& $ 2 $ \\
		\vdots & \vdots & \vdots & \vdots & \vdots & \vdots & \vdots\\
		$(m-1+j)\tau$ & $2(m+j)M $ & $ 2mM $ & $2jM $& $ \mathbf{2mM} $ & $2(m+j-k)(M-1)$ & $2(m+j-k)$ \\
		\vdots & \vdots & \vdots & \vdots & \vdots & \vdots & \vdots\\
		$\mathbf{(k+m-1)\tau}$ & $2(k+m)M $ & $ 2(k+m-j)M $ & $2jM$ & $ 2(k+m-j)M $ &  $ 2m(M-1) $& $ \mathbf{2m} $ \\
		\bottomrule
	\end{tabular}
	\caption{Timing chart for Table~\ref{tab:Rates_Ion_Requirements} Type A, i.e., when $ T\geq\tau_o>\tau_g $, and Table~\ref{tab:Rates_Ion_Requirements} B Case 1, i.e., $(\tau_o>T\geq\tau_g)\wedge (T+\tau_g>\tau_o>T)$. Both of these involve sub-cases 1 and 2 corresponding to $k-j+1\geq m$ and $k-j+1<m$, respectively, where $T=k\tau, \tau_g=j\tau$. The timing diagrams for the different cases of this protocol type are shown in Figs.~\ref{fig:timing_typeAc1} and~\ref{fig:timing_typeAc2}}
	\label{tab:Table_A}
\end{table}

\begin{table}[ht]
	\centering
	\begin{tabular}{c c c c c c c}
		Case 2 & \multicolumn{3}{c}{$Ba^+$ Occupancy} & \multicolumn{2}{c}{$Yb^+$ Occupancy} & \multirow{2}{3cm}{\centering Max. number of heralded ions}
		\\\cmidrule(lr){1-1}\cmidrule(lr){2-4}\cmidrule(lr){5-6}
		Time& Initialized & Freed  & Loaded  & Loaded & Freed &\\
		\midrule\\
		0 & $2M $ & - & $2M $& - & - & - \\
		$\tau$ & $4M $ & - & $4M $ & - & - & - \\
		\vdots & \vdots & \vdots & \vdots & \vdots & \vdots & \vdots\\
		$ k\tau $ & $2(k+1)M $ & $ 2(M-1) $& $2(kM+1) $& - & - & - \\
		\vdots & \vdots & \vdots & \vdots & \vdots &\vdots & \vdots\\
		$(k+j)\tau$ & $2(k+j+1)M $ & $ 2(M-1)(j+1)+2$ & $\mathbf{2(kM+j)} $& $ 2$ &  - & $2$\\
		\vdots & \vdots & \vdots & \vdots & \vdots & \vdots & \vdots\\
		$\mathbf{(k+j+m-1)\tau}$ & $2(k+j+m)M $ & $ 2(M-1)(j+m)+2m $ & $2(kM+j) $& $ \mathbf{2m} $ &  $ - $& $ \mathbf{2m} $ \\
		\bottomrule
	\end{tabular}
	\caption{Timing chart for Tables~\ref{tab:Rates_Ion_Requirements} B and C, Case 2, i.e., when $(\tau_o>T\geq\tau_g)\wedge(\tau_o\geq T+\tau_g>T)$ and $ (\tau_o>\tau_g>T)\wedge(\tau_o\geq T+\tau_g>\tau_g)$, respectively.The timing diagram for this protocol type is shown in Fig.~\ref{fig:timing_typeB}.}
	\label{tab:Table_B}
\end{table}

\begin{table}[ht]
	\centering
	\begin{tabular}{c c c c c c c}
		Case 2 & \multicolumn{3}{c}{$Ba^+$ Occupancy} & \multicolumn{2}{c}{$Yb^+$ Occupancy} & \multirow{2}{3cm}{\centering Max. number of heralded ions}
		\\\cmidrule(lr){1-1}\cmidrule(lr){2-4}\cmidrule(lr){5-6}
		Time & Initialized  & Freed  & Loaded & Loaded& Freed &\\
		\midrule\\
		0 & $2M $ & -& - & - & - & - \\
		$\tau$ & $4M $ & -& - & - & - & - \\
		\vdots & \vdots & \vdots & \vdots & \vdots & \vdots & \vdots\\
		$ k\tau $ & $2(k+1)M $ & - & -& - & - & - \\
		\vdots & \vdots & \vdots & \vdots & \vdots & \vdots & \vdots\\
		$j\tau$ & $2(j+1)M $ & $ 2M$ & $ \mathbf{2jM}$ &  $2M$ & $2(M-1)$& 2\\
		\vdots & \vdots & \vdots & \vdots & \vdots & \vdots & \vdots\\
		$\mathbf{(j+m-1)\tau}$ & $2(j+m)M $ & $ 2mM $ & $ 2jM $ &  $ \mathbf{2mM} $& $2m(M-1)$& $ \mathbf{2m} $ \\
		\bottomrule
	\end{tabular}
	\caption{Timing chart for Table~\ref{tab:Rates_Ion_Requirements} C, Case 1, i.e., when $ (\tau_o>\tau_g>T)\wedge(T+\tau_g>\tau_o>\tau_g)$. The timing diagram for this protocol type is shown in Fig.~\ref{fig:timing_typeC}.}
	\label{tab:Table_C}
\end{table}

\section{Optimization Process}
\label{appendix:optimization}
\begin{figure}[h!]
	\centering
	\includegraphics[width=0.8\linewidth]{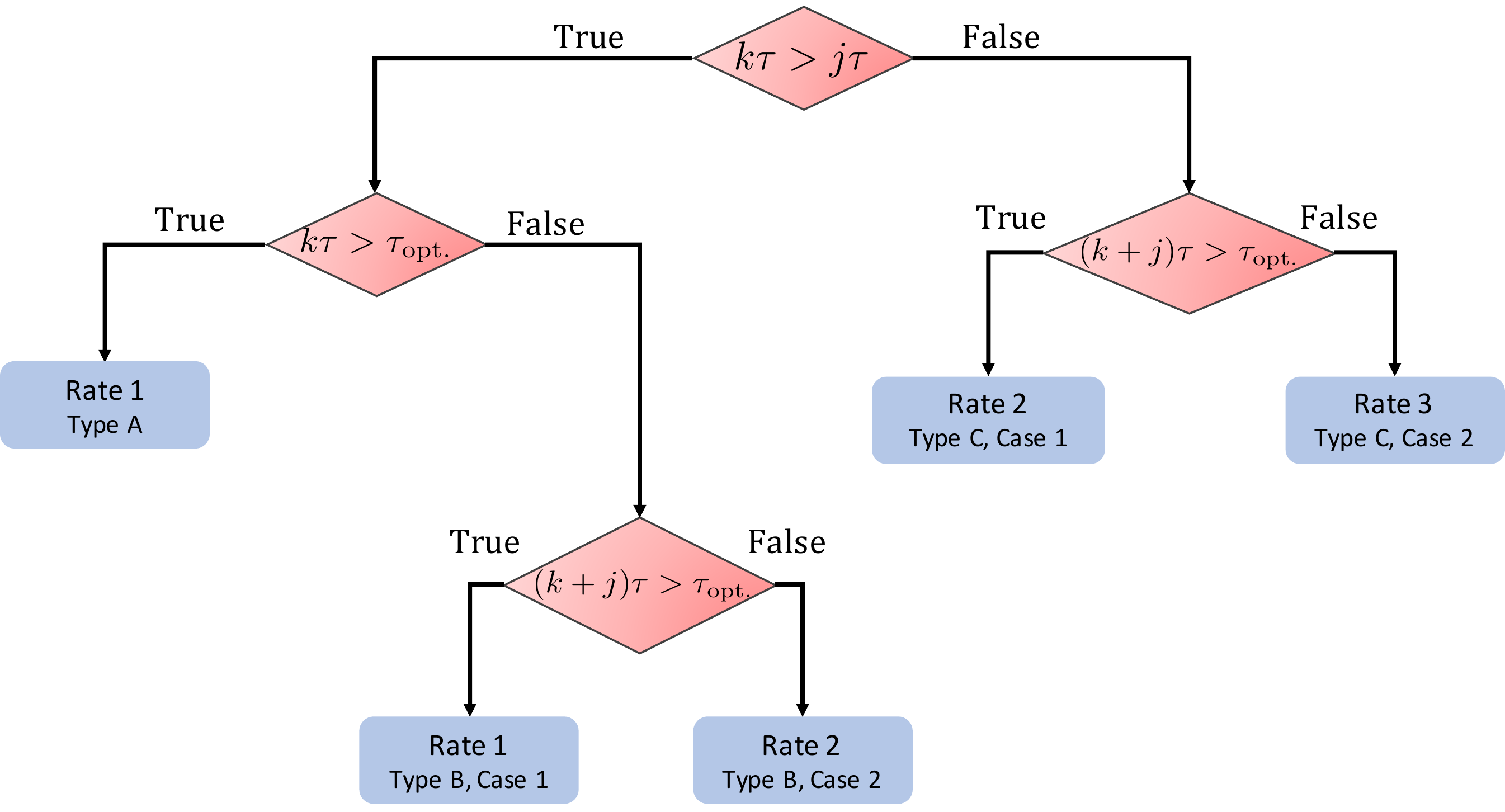}  
	\caption{Decision tree to determine relative timing parameter ordering and associated rate equations.}
	\label{fig:optim}
\end{figure}

Since the protocol in Section~\ref{sec: protocols} is determined by relation between the timing parameters, it is not directly apparent, which rate equation holds true for a given set of protocol parameters (refer Table~\ref{tab:Network_parameters}). The number of repeaters plays a primarily role in determining the heralding time $T$. For the present numerical analysis, we find the optimal parameter values for a given set of conditions using standard optimization techniques. Depending on the optimal values calculated, we now have to make a decision about which type of the rate equation from Table~\ref{tab:Rates_Ion_Requirements} is actually applicable. This is done by traversing the decision tree for the optimal parameter values shown in Fig.~\ref{fig:optim}. One can note that based on the conditions (red diamonds) that are satisfied, the end leaves of the decision tree indicate which rate equation holds true (blue boxes).

\end{document}